\documentclass[aps,prl,reprint,amsmath,amssymb,superscriptaddress,nofootinbib]{revtex4-2}

\usepackage{graphicx}
\usepackage{amsmath}
\usepackage{gensymb}
\usepackage{upgreek}
\usepackage{hyperref}
\usepackage[T1]{fontenc}
\usepackage{xcolor}

\begin{document}

\title{Ballast charges for semiconductor spin qubits}

\author{Yujun Choi}\email{yujunchoi@vt.edu}
\affiliation{Department of Physics, Virginia Tech, Blacksburg, VA 24061, United States}
\affiliation{Virginia Tech Center for Quantum Information Science and Engineering, Blacksburg, VA 24061, United States}

\author{John M. Nichol}
\affiliation{Department of Physics and Astronomy, University of Rochester, Rochester, NY 14627, United States}

\author{Edwin Barnes}
\affiliation{Department of Physics, Virginia Tech, Blacksburg, VA 24061, United States}
\affiliation{Virginia Tech Center for Quantum Information Science and Engineering, Blacksburg, VA 24061, United States}

\date{\today} 

\begin{abstract}
Semiconductor spin qubits are an attractive platform for quantum computing, but their performance is degraded primarily by fluctuating electromagnetic environments.  We introduce the concept of ballast charges, which are induced charges on the surface of an additional screening layer situated below the qubits.  The counteractive behavior of these charges can significantly reduce the power spectral density associated with fluctuations from two-level systems that contribute to charge noise.  Our simulations show that the dephasing time of a spin qubit in a Si/SiGe device increases by a factor of 4 to 6 on average when using this method.  We also discuss the physical implementation and potential challenges of this approach.
\end{abstract}

\maketitle

\textit{Introduction} --- Spin qubits in semiconductor quantum dots (QD) are a promising platform for quantum computing \cite{burkard2023semiconductor}. They have reached the fidelities needed for fault-tolerant operation \cite{xue2022quantum, noiri2022fast, mills2022two, weinstein2023universal} and are also compatible with CMOS manufacturing infrastructure \cite{zwerver2022qubits, neyens2024probing}. One of the main bottlenecks for practical quantum computing with spin qubits is pure dephasing, which typically occurs on a time scale that is orders of magnitude shorter than the spin relaxation time \cite{stano2022review}. This dephasing is primarily due to environmental noise, particularly hyperfine noise from nearby nuclear spins of spinful isotopes and charge noise due to random fluctuations of background charges in QD devices \cite{kuhlmann2013charge}. Charge noise, which typically has a $1/f$ power spectrum \cite{Dial:2013p146804,connors2022charge}, is the dominant noise source in silicon devices, where isotopic enrichment can significantly reduce the density of spinful isotopes, mitigating hyperfine noise \cite{yoneda2018quantum, struck2020low}. The physical origin of charge noise is not yet fully understood \cite{paladino20141}, but there is evidence that two-level systems (TLSs) associated with charge traps in an oxide layer or on the surface of a semiconductor layer are responsible \cite{connors2019low,muller2019towards,ye2024characterization}.

In quantum computing, there are many approaches to combatting environmental noise. One of them is to encode logical information into many physical qubits utilizing quantum error correction codes, which is necessary for fault-tolerant quantum computation \cite{devitt2013quantum, roffe2019quantum}. More near-term methods involve quantum error mitigation techniques implemented at the algorithmic level, such as zero-noise extrapolation, probabilistic error cancellation, etc. \cite{endo2018practical, StrikisPRXQuantum2021}. Other near-term error mitigation techniques include dynamical decoupling and dynamically corrected gates, decoherence-free subspaces, active feedforward loops, and tuning to sweet spots in the parameter space of a particular qubit device \cite{suter2016colloquium, lidar2014review, singh2023mid, barnes2022dynamically}.

In this Letter, we introduce an alternative approach to reducing the effect of charge noise in QD spin qubits that makes use of \emph{ballast} charges in the device.  The concept of ballast charges originates from the idea of ballast water used for maritime vessels. In that context, ballast water is used to control the buoyant force acting on vessels to improve their stability and maneuverability in various maritime conditions; for example, the loading of heavy ballast water can provide added stability when bad weather is expected \cite{david2015vessels}. Analogously, our central idea is to use ballast charges to make spin qubits less susceptible to charge noise, leading to an increase in the dephasing time of the qubits.  The underlying principle of ballast charges relies on electrostatic induction, where induced charges are redistributed to counteract the change in electric fields that are external to the charges. Hence, the ballast charges for spin qubits should be a kind of induced charge that can move freely to compensate for the fluctuations of TLSs.

This approach can also be regarded as leveraging the screening effect of additional electrons in the device. A recent experimental study indirectly exhibits the screening effect by adding electrons to a charge sensor close to a QD \cite{paquelet2023reducing}. Other studies show that the coherence of a spin qubit can be enhanced when the qubit is defined with multiple electrons in a QD \cite{barnes2011screening, higginbotham2014coherent}. In such a scheme, however, we can only expect a small screening effect from induction because the electrons are strongly confined within the QD. Furthermore, the extra electrons complicate gate operations due to the inherent complexity (large Hilbert space) of the multi-electron qubit. This suggests that ballast charges should be extrinsic to the QD and to the qubit itself.

One possible way to introduce more mobile ballast charges is to insert a metallic layer below the quantum well (QW) in which the 2D electron gas (2DEG) that is used to form the qubits resides. This is a natural choice in terms of electrostatic induction. Alternatively, an additional 2DEG can also be used for this purpose, as this would also allow electrons to move freely. For now, we suppose that there is an additional layer in the device that can host ballast charges, and the merits and drawbacks of different physical implementations will be discussed later.

  \begin{figure*}[t]
     \centering
     \includegraphics[width=0.9\textwidth]{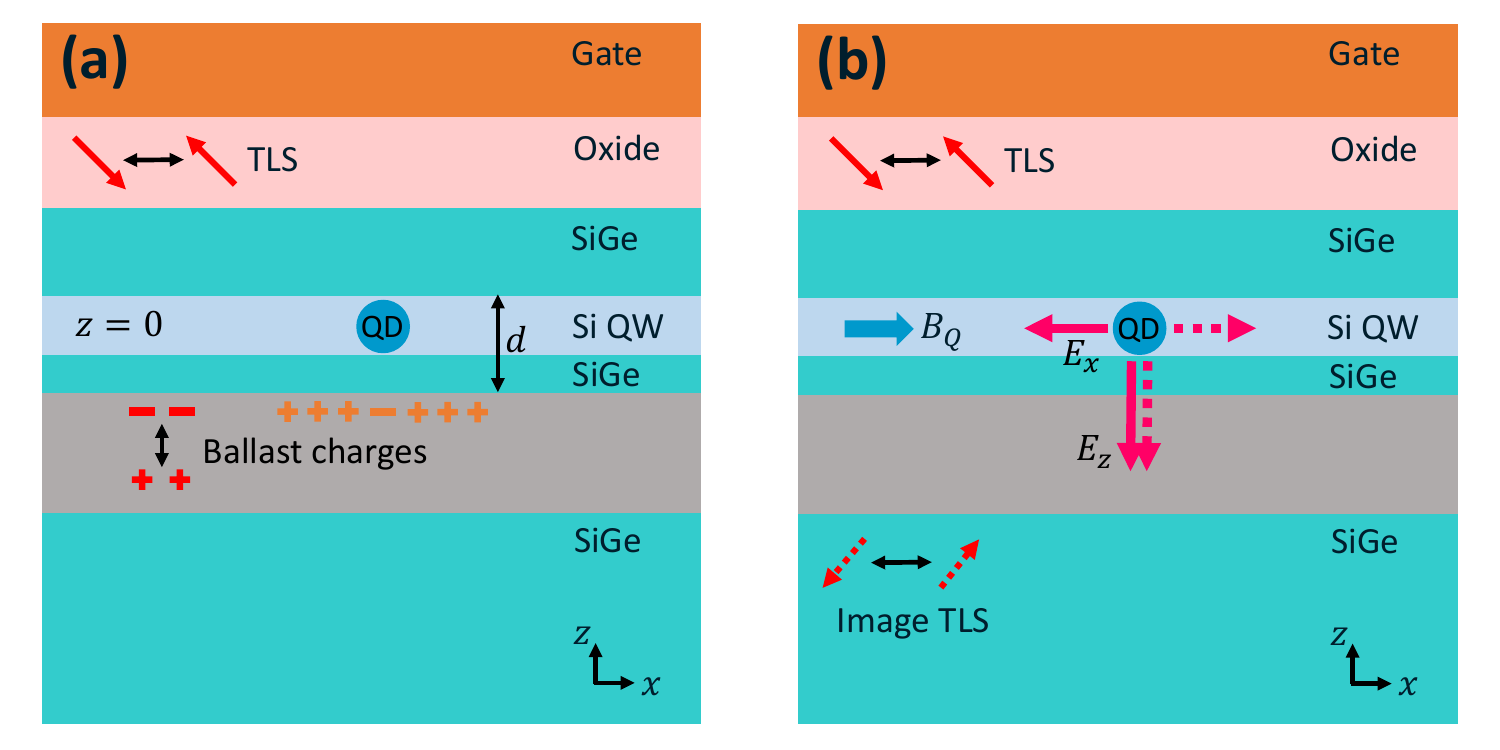}
     \caption{Schematic diagrams for ballast charges in a Si/SiGe quantum dot device. (a) Ballast charges are induced charges due to TLSs (red $\pm$) and gate voltages (orange $\pm$).  The ballast charges are on the surface of the additional screening layer below thin Si quantum well and SiGe layers.  (b) Different viewpoint for ballast charges using the method of image charges. The electric fields ($E_x$ and $E_z$) at the QD position come from TLSs (solid arrows) and their image TLSs (dashed arrows). The diagrams are not to scale.  Si cap between the SiGe and oxide layers and micromagnets on top of the gate layer are not shown.}
     \label{fig:schematic}
 \end{figure*}

A schematic diagram for ballast charges in a Si/SiGe QD device is shown in Fig. \ref{fig:schematic}(a). A screening layer is inserted below a thin SiGe layer that confines a QD in the Si QW. The QD plane is defined as $z=0$, and the distance between the QD plane and the screening layer is $d$. There are two kinds of ballast charges on the surface of the screening layer: One is due to TLSs (red $\pm$) and the other is due to applied gate voltages (orange $\pm$) where $\pm$ denote positive and negative charges on the surface of the screening layer respectively. The ballast charges due to TLSs are the key to the suppression of charge noise since they compensate for the change in electric potential or field at the QD via redistribution. If a TLS is modeled as a flipping electric dipole whose position and orientation are fixed, the ballast charges will change their signs as the TLS flips back and forth. The ballast charges due to gate voltages can be taken into account as offsets when calibrating and tuning the device in the first place.

The role of ballast charges can be understood by the method of image charges as shown in Fig. \ref{fig:schematic}(b). An image TLS corresponding to the TLS in the oxide layer is illustrated as dashed red arrows. Assuming the screening layer is much larger than the size of the QD and TLS and exactly located at the QD plane, the horizontal components of the electric field due to the real and image TLSs perfectly cancel out, while the vertical component is doubled. This gives rise to zero detuning noise provided that there is an in-plane external magnetic field ($B_Q$), and the dominant magnetic field gradients due to micromagnets are $\partial B_Q / \partial x$ and $\partial B_Q / \partial y$. However, this is not what we want, because the qubit cannot be defined due to the equipotential surface along the QD plane. Moreover, it is not possible to make the screening layer at the QD plane in practice. Only partial cancellation of the charge noise will appear when the screening layer is a finite distance away from the QD plane (see Supplemental Material for an analytical result on this).

  \begin{figure}[t]
     \centering
     \includegraphics[width=0.48\textwidth]{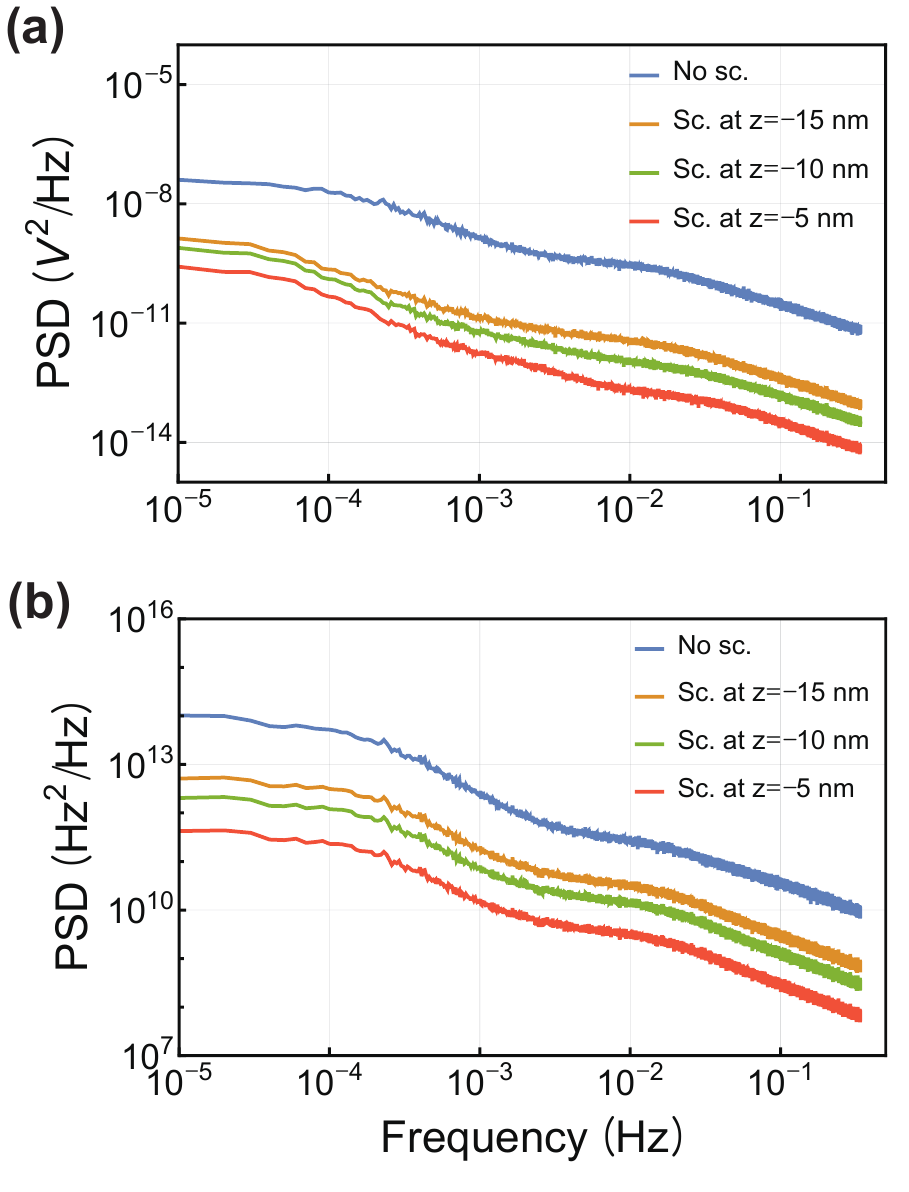}
     \caption{An example of simulated power spectral densities for (a) electric potential and (b) spin qubit frequency for several distances $d$ between the QD and screening layer (Sc. at $z=-d$). The case of no screening layer (No sc.) is also shown.}
     \label{fig:PSD}
 \end{figure}

\textit{Results} --- Examples of simulated power spectral densities (PSDs) are shown for the electric potential and the spin qubit precession frequency in Figs. \ref{fig:PSD}(a) and (b), respectively. TLSs are assumed to be at the interface of the oxide and gate layers ($z=50$ nm) with random orientations. The orders of magnitude of the PSD for the spin qubit frequency agree well with the experimental data in Ref.~\cite{struck2020low}. As expected, the reduction in the PSDs increases as the screening layer is located closer to the QD plane from $z = -15$ to $z = -5$ nm. For a 10 nm Si QW, which is a typical width for realistic devices, the PSD for the electric potential is decreased by about two orders of magnitude. If we make thin Si QW and SiGe layers of 5 nm thickness each \cite{paquelet2023reducing}, the PSD can be reduced by about three orders of magnitude. The reduction in the PSD for the spin qubit frequency is less than that for the electric potential, which is about two orders of magnitude for $d = 5$ nm. This can be attributed to the vector property of compensating electric fields that are more sensitive to the positions and orientations of the image TLSs in the device.

The dephasing time of the qubit can be directly computed from the PSD for the spin qubit frequency assuming Gaussian noise \cite{schriefl2006decoherence}. The dephasing times for no screening, screening at $z=-15$ nm, $z=-10$ nm, and $z=-5$ nm are 2.1, 7.5, 11.7, and 25.2 $\upmu$s, respectively. This enhancement can be understood from the relation $T_2^* \sim 1/\sqrt{\int S(\omega) d\omega}$ in the quasistatic limit \cite{paquelet2023reducing}. The simulation parameters and details are given in the Supplemental Material. This order of magnitude increase in the dephasing time obtained by placing an additional screening layer under thin Si QW and SiGe layers is quite promising.

The example above is one of the TLS realizations in which the screening is very effective ($\sim$ top 1\% of the simulated realizations). The PSD for the spin qubit frequency is typically reduced by one to two orders of magnitude for $d = 5$ nm, and varies a lot in the given frequency range depending on details such as TLS positions, orientations, and switching rates. To capture this variability, we simulate many possible TLS realizations to determine how much the dephasing time can be enhanced on average. Table \ref{tab:Tphi} shows the results for average dephasing time ($T_\phi$) and the increase factor (IF), which is the ratio between the dephasing times with and without screening, for three different cases: Case (A): TLSs located at the interface ($z=50$ nm) with random orientations; Case (B): TLSs in the oxide layer ($30 < z < 50$ nm) with random orientations; and Case (C): TLSs at the interface with vertical orientations (along $\pm z$ directions). In all cases, the TLSs are assumed to have random positions and switching rates, consistent with the random formation of defects in the devices.  The physical models relevant to Case (B) and (C) correspond to the random dipole and trap model respectively, which are discussed in Ref. \cite{choi2022anisotropy}.

The mean values of $T_\phi$ are not so important because it is not valid to compare them between different screening scenarios; they can be used to validate whether the simulation parameters are reasonable or not. In all cases, they are about 1 $\upmu$s without screening, which agrees with values for natural Si devices \cite{stano2022review}. Case (B) has larger mean values than Case (A) since TLSs can be closer to the QD. The standard deviations in parentheses look very large, which is not surprising because the dephasing times have positively skewed distributions (see Supplemental Material). If we change the representative values from mean to median, they decrease in all cases.

A more important quantity is the IF since it is obtained for each realization and can be directly compared with other IFs. Cases (A) and (B) have similar IF values, while those for Case (C) have the largest IFs. This indicates that the screening is most effective for vertically oriented TLSs. The IFs for $d = 10$ nm are about 2-3, corresponding to a slightly less than one order of magnitude reduction in the PSD.  The IFs for $d = 5$ nm are promising, with values ranging from 4 to 6, which means there is a more than one order of magnitude reduction in the PSD.

\begin{table}[t]
\centering
\begin{tabular}{| c | c | c | c | c | c |} 
 \hline
 Case &  & No sc. & $z=-15$ nm & $z=-10$ nm & $z=-5$ nm \\ 
 \hline
 (A) & $T_\phi$ & 1.1 (0.7) & 2.2 (2.2) & 3.1 (3.2) & 5.5 (6.2)\\
 & IF & 1 & 1.8 (0.6) & 2.4 (1.0) & 4.3 (2.1)\\
 \hline
 (B) & $T_\phi$ & 0.9 (0.7) & 1.9 (2.2) & 2.5 (3.1) & 4.7 (6.0)\\
 & IF & 1 & 1.8 (0.7) & 2.4 (1.2) & 4.4 (2.7)\\
 \hline
 (C) & $T_\phi$ & 1.0 (0.7) & 3.0 (5.8) & 4.5 (8.6) & 8.9 (16.3)\\
 & IF & 1 & 2.2 (1.5) & 3.2 (2.6) & 6.1 (5.6)\\
 \hline
\end{tabular}
\caption{Mean (standard deviation) of dephasing time ($T_\phi$) in units of $\upmu$s and its dimensionless increase factor (IF). No sc. and $z=-d$ nm denote no screening layer and screening layer at $z=-d$ nm respectively.}
\label{tab:Tphi}
\end{table}

\textit{Physical implementation} --- Although the above results indicate that ballast charges are effective at screening charge noise, this of course comes at the cost of additional fabrication steps to implement the screening layer.  For GaAs devices, it is possible to epitaxially grow a metallic layer such as Al~\cite{ashlea2021high} to serve this purpose. In-situ growth of epitaxial metal layers may also be possible for silicon devices~\cite{cheng2016epitaxial}. Epitaxial layers are likely to be highly effective sources of ballast charges. However, creating such layers would require significant fabrication development and patterning of such layers will likely be challenging.  However, there are a few other options to consider.

One promising option involves using double quantum well structures. In this structure, the lower electron (or hole) gas serves as a screening layer for the upper 2DEG. Such structures have been fabricated and operated in GaAs/AlGaAs~\cite{nakamura2019aharonov}, Si/SiGe~\cite{Borselli:2011p123118}, and Ge/SiGe platforms~\cite{ivlev2024coupled}. Similarly, adding a Ge spike to a Si QW could be worth trying to form the structure in a single Si layer \cite{mcjunkin2021valley}. It may also be possible to create a double-well structure for SiMOS devices by using silicon-on-insulator substrates, together with a doped silicon substrate that could function as a back gate. 

Another promising technique, which is especially relevant for donor-qubit devices, is scanning tunneling microscope lithography. By using this technique, one can create degenerately doped regions that function like metallic areas.  One could then regrow Si over those areas and then pattern the qubit and gates on top. In principle, this technique could also be applied to other silicon-based spin-qubit systems like Si/SiGe and SiMOS qubits. 

A final approach could be to use ion implantation and epitaxial regrowth to create doped metallic regions inside the semiconductor for silicon spin qubits. Challenges with this approach involve minimizing crystal damage due to the implantation.

\textit{Discussion} --- The principle of ballast water is to increase the inertia of a vessel to fight against external fluctuations (or similarly imagine that the vessel is changed from a small boat to a large cruise ship). This is at the expense of controllability over the vessel. Ballast charges act in the same way, making spin qubits more stable against charge fluctuations while requiring more energy (or gate voltage) to perform gate operations on the qubits. This is because ballast charges will also work to counteract these intentional changes to gate voltages. Just as ballast water can be discharged or loaded depending on what the circumstances require, here too it might be beneficial to turn on and off the function of ballast charges during a quantum circuit. For example, ballast charges could be turned on (off) when the qubits are idle (undergoing gate operations). This would call for switchable discrete patches rather than the always-on monolithic layer considered above. Such patches would be particularly useful for modular quantum dot systems \cite{vandersypen2017interfacing, Li:2018p7}, where each module is mainly affected by nearby local TLSs \cite{rojas2023spatial}.

Another issue related to the controllability is weakened confining potential at the QD plane.  In the case of a grounded screening layer, for example, a confining potential, $V_0$, becomes roughly $V_0 d/(l+d)$ where $l$ ($d$) is the distance between the QD plane and the gate layer (screening layer). With $l = 50$ nm and $d = 5$ nm, the potential becomes $V_0/11$, requiring 11 times larger gate voltage for sustaining the original potential value. This may cause dielectric breakdown of overlapping gates. A patterned screening layer that exactly mirrors the potential by applying an offset voltage, $-V_0 l/d$, to the layer can be used, but patterning without introducing more defects is a big technical challenge. One practical solution for this is to adopt an unpatterned, uniform layer with a non-overlapping-gate architecture~\cite{ha2021flexible,zwerver2022qubits}. Such architectures are more robust against dielectric breakdown because the inter-gate dielectric layers are much thicker than in overlapping-gate devices.

As mentioned earlier, adding the screening layer should not incur additional TLSs near the qubits. The effects of such TLSs would be more critical for the dephasing of the qubits than the remote TLSs in the oxide layer because of their short distances from the qubits. Although the horizontal component of the TLSs will be screened by the layer, the vertical one will be antiscreened, leading to larger fluctuations.  The screening layer will thus be effective when the dominant charge noise sources are the remote TLSs. In addition, most studies have suggested the charge noise originates from interfaces or materials above the surface of the semiconductor.

The qubits in the vicinity of a metallic layer can be affected by evanescent-wave Johnson noise (EWJN) \cite{langsjoen2014}. Electric fields due to EWJN would have a larger impact on the qubits than magnetic fields, because the magnitude of the former is proportional to $1/d^3$, while the latter is proportional to $w/d^2$ \cite{premakumar2017evanescent}. Placing the layer close to the qubits results in an increased contribution from EWJN; it is estimated to be $T_\phi^{EWJN} = 4.7$ ms for $d=5$ nm with reasonable thickness $w = 30$ nm and conductivity $\sigma = 1.6\times 10^7$ S/m \cite{Tenberg2019}. The relaxation time due to EWJN is about $T_1^{EWJN} = 27$ ms, which is small compared to the measured $T_1$ of conventional Si devices, and so EWJN could become the main source of relaxation. For $T_\phi^{EWJN}$, the electric part is one order of magnitude larger than the magnetic part, while only the magnetic part contributes to $T_1^{EWJN}$.

Heating effects from the metallic layer should also be considered, because they can shift the spin qubit frequency during gate operations \cite{undseth2023hotter, choi2024interacting}. Increasing the conductivity of the layer can reduce the heat from dissipation. Since the magnetic and electric parts of EWJN are proportional to $\sigma$ and $1/\sigma$, respectively \cite{choi2022anisotropy}, at some point the magnetic part will exceed the electric part and become the dominant dephasing noise source.

A superconducting thin film may be able to serve as the layer to minimize the heat from dissipation.  Its screening effect would be comparable to a normal metal \cite{Amoretti_2023, zaccone2024theory}, but this needs careful consideration because EWJN may play a significant role in dephasing. The superconductivity in a Ge QW induced by a proximity effect could also be employed~\cite{tosato2023hard}.

The redistribution of charges itself can cause some noise, but the relevant time scale is less than picoseconds. This can be ignored because this noise contributes to the PSD at frequencies beyond the THz range, and so the effect on $T_\phi$ and $T_1$ is limited.

An important question is how close to the qubits can the ballast charges be placed. If they are too close, their wave functions will overlap with those of the spin qubits, and unwanted interactions may occur. Since the ballast charges would be numerous and interact with one another, any unwanted entanglement with the qubits through the charge degree of freedom is expected to decay fast \cite{mohanty2002decoherent}. Indirect interactions between the qubits mediated by the ballast charges, such as Ruderman-Kittel-Kasuya-Yosida interactions, may need to be accounted for to achieve precise gate operations in this regime \cite{yang2016long, tanamoto2021compact}.

\textit{Summary and outlook} --- We have introduced the idea of ballast charges for semiconductor spin qubits. This amounts to placing an additional screening layer below the QD plane where spin qubits reside. We have shown that the PSD for the spin qubit precession frequency can be reduced by 1-2 orders of magnitude in a Si/SiGe device. Through simulations over many TLS realizations, we have confirmed that the reduction gives rise to a 4 to 6-fold increase in dephasing times on average for thin Si QW and SiGe layers. We have discussed the physical implementation of the layer in various semiconductor devices. We have also explored other features of ballast charges and potential issues.

A major benefit of using ballast charges is their passive character. This approach does not require complicated control protocols or computational resources for active feedforward. Therefore, ballast charges are compatible with other techniques such as quantum error correction codes and error mitigation. It would also be intriguing to apply this idea to other qubit platforms as well. While physical realizations may vary, the idea of exploiting the innate property of an additional system to passively counteract external fluctuations could have wide applicability.

\section{Acknowledgments}

JMN acknowledges funding from the Army Research Office through Grant No. W911NF-23-1-0115 and the Air Force Office of Scientific Research through Grant No. FA9550-23-1-0710. EB acknowledges support from the Army Research Office through Grant No. W911NF-23-1-0115.

\onecolumngrid

\vspace{20mm}

\section{Supplemental material}

\subsection{Analytic calculations using the method of image charges}

Suppose that a spin qubit is located at the origin of the coordinate system, and a two-level system (TLS) is located at $\vec{r} = (x, y, z)$ with $z>0$ with dipole vector $\vec{p} = (p_x, p_y, p_z)$. The electric potential at the qubit position due to the TLS is
\begin{equation}
	V_{\mathrm{dip}} = \frac{-\vec{p}\cdot\vec{r}}{4\pi\epsilon_r \epsilon_0 |\vec{r}|^3}, 
\end{equation}
where $\epsilon_r$ is the dielectric constant, and $\epsilon_0$ is the vacuum permittivity.  The screening layer is assumed to be at $z=-d$ ($d>0$), and so the position and the dipole vector of its image TLS are $\vec{r}_{\mathrm{im}} = (x, y, -z-2d)$ and $\vec{p}_{\mathrm{im}} = (-p_x, -p_y, p_z)$. The potential at the qubit position is the sum of the potentials of the real and image TLSs:
\begin{align}
	V_{\mathrm{sc}} &= V_{\mathrm{dip}} + V_{\mathrm{im}} = \frac{1}{4\pi\epsilon_r \epsilon_0} \left( \frac{-\vec{p}\cdot\vec{r}}{|\vec{r}|^3} + \frac{-\vec{p}_{\mathrm{im}}\cdot\vec{r}_{\mathrm{im}}}{|\vec{r}_{\mathrm{im}}|^3} \right) \\
 &= \frac{1}{4\pi\epsilon_r \epsilon_0} \left( \frac{-p_x x - p_y y - p_z z}{(x^2 + y^2 + z^2)^{3/2}} + \frac{p_x x + p_y y + p_z (z+2d)}{(x^2 + y^2 + (z+2d)^2)^{3/2}} \right).
\end{align}
When $d = 0$, they perfectly cancel out so the potential $V_{\mathrm{sc}}$ is exactly zero.

The electric field at the qubit position due to the TLS is
\begin{equation}
    \vec{E}_{\mathrm{dip}} =  \frac{3 (\vec{p} \cdot \vec{r} ) \vec{r} - |\vec{r}|^2\vec{p}}{4 \pi \epsilon_r \epsilon_0 |\vec{r}|^5} \equiv (E_{\mathrm{dip},x}, E_{\mathrm{dip},y}, E_{\mathrm{dip},z}).
\end{equation}
The electric field at the qubit position is also the sum of the fields from the real and image TLSs:
\begin{equation}
\vec{E}_{\mathrm{sc}} = \vec{E}_{\mathrm{dip}} + \vec{E}_{\mathrm{im}} = \frac{1}{4\pi\epsilon_r \epsilon_0} \left( \frac{3 (\vec{p} \cdot \vec{r} ) \vec{r} - |\vec{r}|^2\vec{p}}{|\vec{r}|^5} + \frac{3 (\vec{p}_{\mathrm{im}} \cdot \vec{r}_{\mathrm{im}} ) \vec{r}_{\mathrm{im}} - |\vec{r}_{\mathrm{im}}|^2\vec{p}_{\mathrm{im}}}{|\vec{r}_{\mathrm{im}}|^5} \right) \equiv (E_{\mathrm{sc},x}, E_{\mathrm{sc},y}, E_{\mathrm{sc},z}). 
\end{equation}
When $d = 0$, $E_{\mathrm{sc},x} = E_{\mathrm{sc},y} = 0$ and $E_{\mathrm{sc},z} = 2 E_{\mathrm{dip},z}$.

\subsection{Simulation parameters}

The position of a spin qubit is $(x,y,z) = (0, 0, 0)$ nm. TLS positions are randomly sampled from the uniform distribution over $-150 < x,y < 150$ nm with $z = 50$ nm for Case (A) and (C), and with $30 < z < 50$ nm for Case (B). Since 10 TLSs are used for each realization, the number density corresponds to $1.11\times 10^{10} \,\mathrm{cm}^{-2}$ for Case (A) and (C) \cite{kkepa2023correlations, kkepa2023simulation}, and $5.56\times 10^{15} \,\mathrm{cm}^{-3}$ for Case (B), respectively. The switching rates of TLSs are log-uniformly sampled from $\nu \in (10^{-5}, 1)$ Hz to ensure their power spectral density has a $1/f$-like form. The dielectric constant $\epsilon_r$ used in the simulation is 11. The dipole magnitude is set to $p_0 = 0.5 \,|e|\cdot$nm $\approx 24$ Debye, which is a representative value with elementary charge $e$ hopping between two locations separated by a distance of 0.5 nm.  

For the spin qubit precession frequency, we need to clarify the artificial spin-orbit coupling introduced by magnetic field gradients of the micromagnets on top of the Si/SiGe device. The gradients convert electric fields to effective magnetic fields at the qubit position. We follow the same convention used in Ref. \cite{yoneda2023noise}: The change in the spin qubit frequency (in the units of Hz) is
\begin{equation}
    \delta f_{\mathrm{spin}} = \lambda_Q \delta E_{x},
\end{equation}
where $\delta E_{x}$ is the change in the $x$ component of the electric field at the qubit position and $\lambda_Q = \gamma_e (\partial B_Q / \partial x) \kappa_Q$ with electron gyromagnetic ratio $\gamma_e = -28$ GHz/T, magnetic field gradient $\partial B_Q / \partial x = 0.3$ mT/nm, and a factor $\kappa_Q = e/(m \omega_{\mathrm{orb}}^2)$. The factor $\kappa_Q$ is related to the orbital confinement energy $\hbar \omega_{\mathrm{orb}}$, which is estimated to be 1 meV. The effective electron mass is set to $m = 0.2 m_e$ with electron rest mass $m_e$. Since the change in the electric field is purely due to TLSs for our purpose, $\delta E_x = \sum_k E_{\mathrm{dip},x}^{(k)}$ ($\delta E_x = \sum_k E_{\mathrm{sc},x}^{(k)}$) without (with) screening with TLS index $k$.

\subsection{Simulation details}

\begin{figure*}
    \centering   \includegraphics[width=0.96\textwidth]{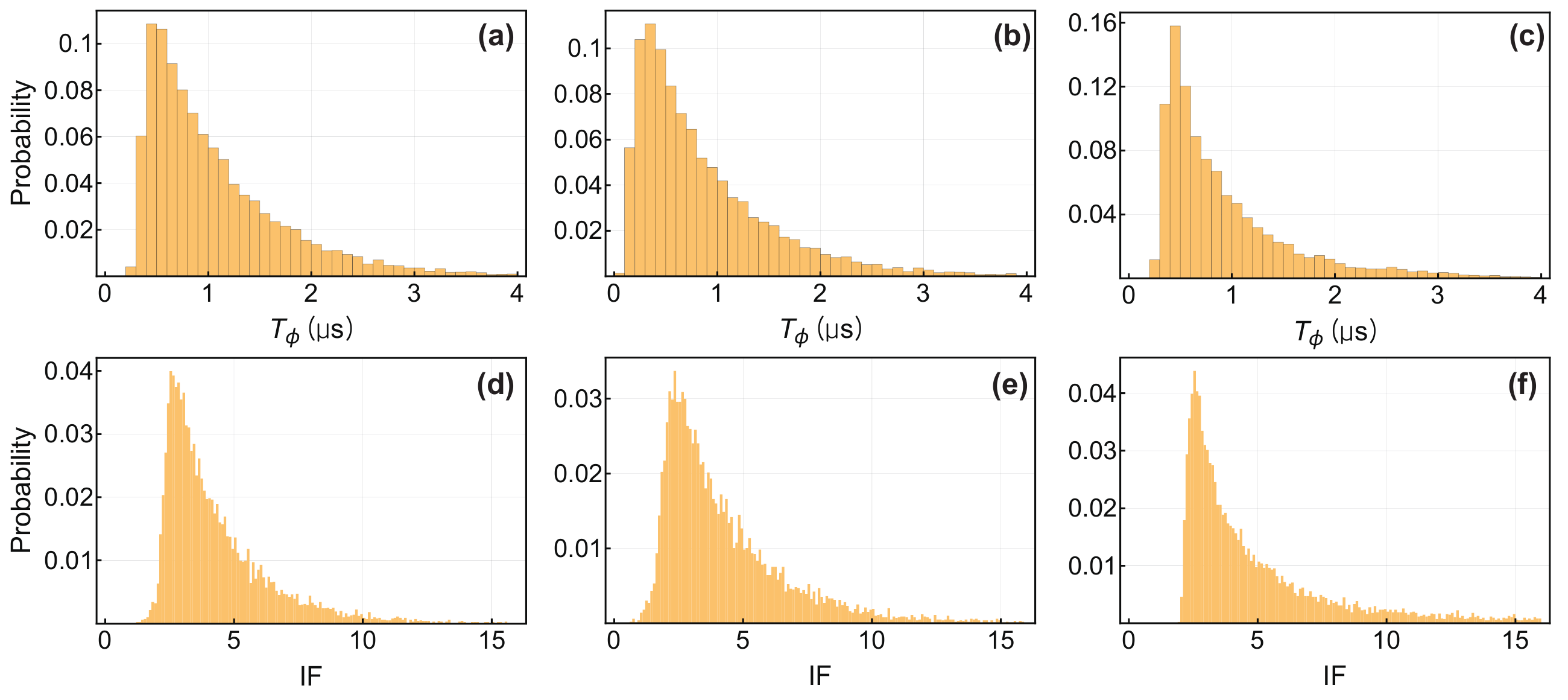}
     \caption{Probability histograms of dephasing time ($T_\phi$) and increase factor (IF).  (a), (b), and (c) show histograms of dephasing time for Case (A), (B), and (C), respectively. (d), (e), and (f) show IFs with $d = 5$ nm for Case (A), (B), and (C), respectively.  The full ranges of $T_\phi$ and IF are larger than those shown in the histograms.}
     \label{fig:histogram}
\end{figure*}

For each case of (A), (B), and (C), we generated $10^4$ realizations with the parameters above.  The power spectral density (PSD) for the spin qubit frequency is defined as
\begin{equation}
    S(f) \equiv \int_{-\infty}^{\infty} \langle \delta f_{\mathrm{spin}}(t) \delta f_{\mathrm{spin}}(t+\tau) \rangle e^{-i 2\pi f \tau} d\tau = \lambda_Q^2 \sum_k E_k^2 \frac{\nu_k}{(\pi f)^2 + \nu_k^2},
\end{equation}
where $\langle \cdot \rangle$ is the statistical average of the correlation function for $\delta f_{\mathrm{spin}}$, $E_k$ is either $E_{\mathrm{dip},x}^{(k)}$ or $E_{\mathrm{sc},x}^{(k)}$ depending on the existence of the screening layer, and the frequency dependence has a Lorentzian form for random telegraph noise \cite{machlup1954noise}. It is assumed that the TLSs do not interact with each other. Given the analytical expression for $S(f)$ above, the dephasing time ($T_\phi$) can be computed assuming Gaussian noise \cite{schriefl2006decoherence}. The attenuation function for free induction decay is 
\begin{equation}
    \chi (t) = 2\pi \frac{t^2}{2} \int_{-\infty}^{\infty} df \, S(f) \, \mathrm{sinc}^2 (\pi f t),
\end{equation}
where $\mathrm{sinc} (x) \equiv \sin (x) / x$.  $T_\phi$ can be obtained numerically by imposing $\chi (T_\phi) = 1$.

The histograms of dephasing time and increase factor (IF) for $10^4$ realizations are shown in Fig. \ref{fig:histogram}. Figs. \ref{fig:histogram}(a)--(c) correspond to $T_\phi$ without a screening layer for Cases (A)--(C), respectively. The IFs in Figs.~\ref{fig:histogram}(d)--(f) are the results with a screening layer 5 nm below Si quantum well ($d=5$ nm) for Cases (A)--(C), respectively.  Most of the dephasing times and IFs are less than 4 $\upmu$s and 15, respectively. Both quantities have skewed distributions. This may be attributed to the sparseness of the bath of TLSs \cite{mehmandoost2024decoherence}. In our simulations, the spectral density of TLSs is $N_{\mathrm{TLS}}/\ln(\nu_M/\nu_m) = 0.87$, where $N_{\mathrm{TLS}} = 10$ is the number of TLSs and $\nu_M = 1$ Hz ($\nu_m = 10^{-5}$ Hz) is the maximum (minimum) switching rate that the TLSs can have. Since the Gaussian approximation is valid when the spectral density is comparable to or larger than 1, $T_\phi$ could be a little bit off from the exact value. Nevertheless, it should suffice to assume Gaussian noise because the purpose of our work is not to obtain $T_\phi$ accurately, but to confirm the increase in $T_\phi$ due to the screening effect. We may be able to increase $N_{\mathrm{TLS}}$ while decreasing $p_0$ for more accurate $T_\phi$ with the Gaussian assumption. This will not be the case when a couple of TLSs dominate the charge noise \cite{connors2019low, rojas2023spatial, ye2024characterization}.

\begin{table}[t]
\centering
\begin{tabular}{| c | c | c | c | c | c |} 
 \hline
 Case &  & No sc. & $z=-15$ nm & $z=-10$ nm & $z=-5$ nm \\ 
 \hline
 (A) & $T_\phi$ & 0.9 (0.8) & 1.5 (1.9) & 1.9 (2.7) & 3.2 (5.1)\\
 & IF & 1 & 1.7 (0.6) & 2.1 (1.0) & 3.7 (2.3)\\
 \hline
 (B) & $T_\phi$ & 0.7 (0.8) & 1.1 (1.8) & 1.4 (2.5) & 2.4 (4.8)\\
 & IF & 1 & 1.6 (0.7) & 2.1 (1.2) & 3.5 (2.7)\\
 \hline
 (C) & $T_\phi$ & 0.7 (0.7) & 1.2 (2.0) & 1.6 (2.9) & 2.8 (5.9)\\
 & IF & 1 & 1.7 (0.9) & 2.3 (1.6) & 4.0 (3.8)\\
 \hline
\end{tabular}
\caption{Median (interquartile range) of dephasing time and its increase factor. The dephasing time ($T_\phi$) is in the units of microseconds, and the increase factor (IF) is dimensionless. No sc. and $z=-d$ nm denote no screening layer and screening layer at $z=-d$ nm, respectively.}
\label{tab:Tphi_median}
\end{table}

The median (interquartile range) values of dephasing times and IFs are shown in Table~\ref{tab:Tphi_median}. They are smaller than the corresponding mean (standard deviation) due to the skewed distribution. The dephasing time can still be increased by a factor of 3-4 for thin Si QW and SiGe layers.

The dephasing times of the example in the main text were computed using the attenuation function given above. The PSDs of the example were produced by averaging over 100 time series generated by Poisson processes with sampled switching rates.

\subsection{Evanescent-wave Johnson noise from metallic screening layer}

The dephasing time and relaxation time due to evanescent-wave Johnson noise (EWJN) were calculated based on the formulas in Ref. \cite{choi2022anisotropy}. 
 One needs to be cautious about the presence of some typos that should be corrected: $1/2k_j \,\to\, 1/k_j$ in Eq. (7), $1/\sqrt{4\pi\epsilon_0} \,\to\, 1/4\pi\epsilon_0$ in Eq. (10), and the equations that are relevant to Eq. (7) and (10). We assume that the external magnetic field is applied along the $x$ direction, so the relevant coherence times are $T_\phi^{(x)}$ and $T_1^{(x)}$ in Ref. \cite{choi2022anisotropy}. They change little even if the magnetic field direction is along the $y$ direction or other in-plane directions.

The simulation parameters given above and in the main text were used to obtain the coherence times due to EWJN. Other parameters needed for the simulation are the qubit operating frequency $\omega_{op} = 2\pi \times 16.3 \times 10^{9} \,\mathrm{s}^{-1}$ and the electron temperature $T = 50$ mK \cite{yoneda2023noise}. If other magnetic field gradients are not negligible as in Ref. \cite{Kawakami:2014p666}, the relaxation time due to EWJN can also be affected by the electric part, reducing it to a few milliseconds.


\vspace{20mm}
\twocolumngrid
\bibliography{main}

\begin{thebibliography}{58}%
\makeatletter
\providecommand \@ifxundefined [1]{%
 \@ifx{#1\undefined}
}%
\providecommand \@ifnum [1]{%
 \ifnum #1\expandafter \@firstoftwo
 \else \expandafter \@secondoftwo
 \fi
}%
\providecommand \@ifx [1]{%
 \ifx #1\expandafter \@firstoftwo
 \else \expandafter \@secondoftwo
 \fi
}%
\providecommand \natexlab [1]{#1}%
\providecommand \enquote  [1]{``#1''}%
\providecommand \bibnamefont  [1]{#1}%
\providecommand \bibfnamefont [1]{#1}%
\providecommand \citenamefont [1]{#1}%
\providecommand \href@noop [0]{\@secondoftwo}%
\providecommand \href [0]{\begingroup \@sanitize@url \@href}%
\providecommand \@href[1]{\@@startlink{#1}\@@href}%
\providecommand \@@href[1]{\endgroup#1\@@endlink}%
\providecommand \@sanitize@url [0]{\catcode `\\12\catcode `\$12\catcode `\&12\catcode `\#12\catcode `\^12\catcode `\_12\catcode `\%12\relax}%
\providecommand \@@startlink[1]{}%
\providecommand \@@endlink[0]{}%
\providecommand \url  [0]{\begingroup\@sanitize@url \@url }%
\providecommand \@url [1]{\endgroup\@href {#1}{\urlprefix }}%
\providecommand \urlprefix  [0]{URL }%
\providecommand \Eprint [0]{\href }%
\providecommand \doibase [0]{https://doi.org/}%
\providecommand \selectlanguage [0]{\@gobble}%
\providecommand \bibinfo  [0]{\@secondoftwo}%
\providecommand \bibfield  [0]{\@secondoftwo}%
\providecommand \translation [1]{[#1]}%
\providecommand \BibitemOpen [0]{}%
\providecommand \bibitemStop [0]{}%
\providecommand \bibitemNoStop [0]{.\EOS\space}%
\providecommand \EOS [0]{\spacefactor3000\relax}%
\providecommand \BibitemShut  [1]{\csname bibitem#1\endcsname}%
\let\auto@bib@innerbib\@empty
\bibitem [{\citenamefont {Burkard}\ \emph {et~al.}(2023)\citenamefont {Burkard}, \citenamefont {Ladd}, \citenamefont {Pan}, \citenamefont {Nichol},\ and\ \citenamefont {Petta}}]{burkard2023semiconductor}%
  \BibitemOpen
  \bibfield  {author} {\bibinfo {author} {\bibfnamefont {G.}~\bibnamefont {Burkard}}, \bibinfo {author} {\bibfnamefont {T.~D.}\ \bibnamefont {Ladd}}, \bibinfo {author} {\bibfnamefont {A.}~\bibnamefont {Pan}}, \bibinfo {author} {\bibfnamefont {J.~M.}\ \bibnamefont {Nichol}},\ and\ \bibinfo {author} {\bibfnamefont {J.~R.}\ \bibnamefont {Petta}},\ }\bibfield  {title} {\bibinfo {title} {Semiconductor spin qubits},\ }\href@noop {} {\bibfield  {journal} {\bibinfo  {journal} {Reviews of Modern Physics}\ }\textbf {\bibinfo {volume} {95}},\ \bibinfo {pages} {025003} (\bibinfo {year} {2023})}\BibitemShut {NoStop}%
\bibitem [{\citenamefont {Xue}\ \emph {et~al.}(2022)\citenamefont {Xue}, \citenamefont {Russ}, \citenamefont {Samkharadze}, \citenamefont {Undseth}, \citenamefont {Sammak}, \citenamefont {Scappucci},\ and\ \citenamefont {Vandersypen}}]{xue2022quantum}%
  \BibitemOpen
  \bibfield  {author} {\bibinfo {author} {\bibfnamefont {X.}~\bibnamefont {Xue}}, \bibinfo {author} {\bibfnamefont {M.}~\bibnamefont {Russ}}, \bibinfo {author} {\bibfnamefont {N.}~\bibnamefont {Samkharadze}}, \bibinfo {author} {\bibfnamefont {B.}~\bibnamefont {Undseth}}, \bibinfo {author} {\bibfnamefont {A.}~\bibnamefont {Sammak}}, \bibinfo {author} {\bibfnamefont {G.}~\bibnamefont {Scappucci}},\ and\ \bibinfo {author} {\bibfnamefont {L.~M.}\ \bibnamefont {Vandersypen}},\ }\bibfield  {title} {\bibinfo {title} {Quantum logic with spin qubits crossing the surface code threshold},\ }\href@noop {} {\bibfield  {journal} {\bibinfo  {journal} {Nature}\ }\textbf {\bibinfo {volume} {601}},\ \bibinfo {pages} {343} (\bibinfo {year} {2022})}\BibitemShut {NoStop}%
\bibitem [{\citenamefont {Noiri}\ \emph {et~al.}(2022)\citenamefont {Noiri}, \citenamefont {Takeda}, \citenamefont {Nakajima}, \citenamefont {Kobayashi}, \citenamefont {Sammak}, \citenamefont {Scappucci},\ and\ \citenamefont {Tarucha}}]{noiri2022fast}%
  \BibitemOpen
  \bibfield  {author} {\bibinfo {author} {\bibfnamefont {A.}~\bibnamefont {Noiri}}, \bibinfo {author} {\bibfnamefont {K.}~\bibnamefont {Takeda}}, \bibinfo {author} {\bibfnamefont {T.}~\bibnamefont {Nakajima}}, \bibinfo {author} {\bibfnamefont {T.}~\bibnamefont {Kobayashi}}, \bibinfo {author} {\bibfnamefont {A.}~\bibnamefont {Sammak}}, \bibinfo {author} {\bibfnamefont {G.}~\bibnamefont {Scappucci}},\ and\ \bibinfo {author} {\bibfnamefont {S.}~\bibnamefont {Tarucha}},\ }\bibfield  {title} {\bibinfo {title} {Fast universal quantum gate above the fault-tolerance threshold in silicon},\ }\href@noop {} {\bibfield  {journal} {\bibinfo  {journal} {Nature}\ }\textbf {\bibinfo {volume} {601}},\ \bibinfo {pages} {338} (\bibinfo {year} {2022})}\BibitemShut {NoStop}%
\bibitem [{\citenamefont {Mills}\ \emph {et~al.}(2022)\citenamefont {Mills}, \citenamefont {Guinn}, \citenamefont {Gullans}, \citenamefont {Sigillito}, \citenamefont {Feldman}, \citenamefont {Nielsen},\ and\ \citenamefont {Petta}}]{mills2022two}%
  \BibitemOpen
  \bibfield  {author} {\bibinfo {author} {\bibfnamefont {A.~R.}\ \bibnamefont {Mills}}, \bibinfo {author} {\bibfnamefont {C.~R.}\ \bibnamefont {Guinn}}, \bibinfo {author} {\bibfnamefont {M.~J.}\ \bibnamefont {Gullans}}, \bibinfo {author} {\bibfnamefont {A.~J.}\ \bibnamefont {Sigillito}}, \bibinfo {author} {\bibfnamefont {M.~M.}\ \bibnamefont {Feldman}}, \bibinfo {author} {\bibfnamefont {E.}~\bibnamefont {Nielsen}},\ and\ \bibinfo {author} {\bibfnamefont {J.~R.}\ \bibnamefont {Petta}},\ }\bibfield  {title} {\bibinfo {title} {Two-qubit silicon quantum processor with operation fidelity exceeding 99\%},\ }\href@noop {} {\bibfield  {journal} {\bibinfo  {journal} {Science Advances}\ }\textbf {\bibinfo {volume} {8}},\ \bibinfo {pages} {eabn5130} (\bibinfo {year} {2022})}\BibitemShut {NoStop}%
\bibitem [{\citenamefont {Weinstein}\ \emph {et~al.}(2023)\citenamefont {Weinstein}, \citenamefont {Reed}, \citenamefont {Jones}, \citenamefont {Andrews}, \citenamefont {Barnes}, \citenamefont {Blumoff}, \citenamefont {Euliss}, \citenamefont {Eng}, \citenamefont {Fong}, \citenamefont {Ha} \emph {et~al.}}]{weinstein2023universal}%
  \BibitemOpen
  \bibfield  {author} {\bibinfo {author} {\bibfnamefont {A.~J.}\ \bibnamefont {Weinstein}}, \bibinfo {author} {\bibfnamefont {M.~D.}\ \bibnamefont {Reed}}, \bibinfo {author} {\bibfnamefont {A.~M.}\ \bibnamefont {Jones}}, \bibinfo {author} {\bibfnamefont {R.~W.}\ \bibnamefont {Andrews}}, \bibinfo {author} {\bibfnamefont {D.}~\bibnamefont {Barnes}}, \bibinfo {author} {\bibfnamefont {J.~Z.}\ \bibnamefont {Blumoff}}, \bibinfo {author} {\bibfnamefont {L.~E.}\ \bibnamefont {Euliss}}, \bibinfo {author} {\bibfnamefont {K.}~\bibnamefont {Eng}}, \bibinfo {author} {\bibfnamefont {B.~H.}\ \bibnamefont {Fong}}, \bibinfo {author} {\bibfnamefont {S.~D.}\ \bibnamefont {Ha}}, \emph {et~al.},\ }\bibfield  {title} {\bibinfo {title} {Universal logic with encoded spin qubits in silicon},\ }\href@noop {} {\bibfield  {journal} {\bibinfo  {journal} {Nature}\ }\textbf {\bibinfo {volume} {615}},\ \bibinfo {pages} {817} (\bibinfo {year} {2023})}\BibitemShut {NoStop}%
\bibitem [{\citenamefont {Zwerver}\ \emph {et~al.}(2022)\citenamefont {Zwerver}, \citenamefont {Kr{\"a}henmann}, \citenamefont {Watson}, \citenamefont {Lampert}, \citenamefont {George}, \citenamefont {Pillarisetty}, \citenamefont {Bojarski}, \citenamefont {Amin}, \citenamefont {Amitonov}, \citenamefont {Boter} \emph {et~al.}}]{zwerver2022qubits}%
  \BibitemOpen
  \bibfield  {author} {\bibinfo {author} {\bibfnamefont {A.}~\bibnamefont {Zwerver}}, \bibinfo {author} {\bibfnamefont {T.}~\bibnamefont {Kr{\"a}henmann}}, \bibinfo {author} {\bibfnamefont {T.}~\bibnamefont {Watson}}, \bibinfo {author} {\bibfnamefont {L.}~\bibnamefont {Lampert}}, \bibinfo {author} {\bibfnamefont {H.~C.}\ \bibnamefont {George}}, \bibinfo {author} {\bibfnamefont {R.}~\bibnamefont {Pillarisetty}}, \bibinfo {author} {\bibfnamefont {S.}~\bibnamefont {Bojarski}}, \bibinfo {author} {\bibfnamefont {P.}~\bibnamefont {Amin}}, \bibinfo {author} {\bibfnamefont {S.}~\bibnamefont {Amitonov}}, \bibinfo {author} {\bibfnamefont {J.}~\bibnamefont {Boter}}, \emph {et~al.},\ }\bibfield  {title} {\bibinfo {title} {Qubits made by advanced semiconductor manufacturing},\ }\href@noop {} {\bibfield  {journal} {\bibinfo  {journal} {Nature Electronics}\ }\textbf {\bibinfo {volume} {5}},\ \bibinfo {pages} {184} (\bibinfo {year} {2022})}\BibitemShut {NoStop}%
\bibitem [{\citenamefont {Neyens}\ \emph {et~al.}(2024)\citenamefont {Neyens}, \citenamefont {Zietz}, \citenamefont {Watson}, \citenamefont {Luthi}, \citenamefont {Nethwewala}, \citenamefont {George}, \citenamefont {Henry}, \citenamefont {Islam}, \citenamefont {Wagner}, \citenamefont {Borjans} \emph {et~al.}}]{neyens2024probing}%
  \BibitemOpen
  \bibfield  {author} {\bibinfo {author} {\bibfnamefont {S.}~\bibnamefont {Neyens}}, \bibinfo {author} {\bibfnamefont {O.~K.}\ \bibnamefont {Zietz}}, \bibinfo {author} {\bibfnamefont {T.~F.}\ \bibnamefont {Watson}}, \bibinfo {author} {\bibfnamefont {F.}~\bibnamefont {Luthi}}, \bibinfo {author} {\bibfnamefont {A.}~\bibnamefont {Nethwewala}}, \bibinfo {author} {\bibfnamefont {H.~C.}\ \bibnamefont {George}}, \bibinfo {author} {\bibfnamefont {E.}~\bibnamefont {Henry}}, \bibinfo {author} {\bibfnamefont {M.}~\bibnamefont {Islam}}, \bibinfo {author} {\bibfnamefont {A.~J.}\ \bibnamefont {Wagner}}, \bibinfo {author} {\bibfnamefont {F.}~\bibnamefont {Borjans}}, \emph {et~al.},\ }\bibfield  {title} {\bibinfo {title} {Probing single electrons across 300-mm spin qubit wafers},\ }\href@noop {} {\bibfield  {journal} {\bibinfo  {journal} {Nature}\ }\textbf {\bibinfo {volume} {629}},\ \bibinfo {pages} {80} (\bibinfo {year} {2024})}\BibitemShut {NoStop}%
\bibitem [{\citenamefont {Stano}\ and\ \citenamefont {Loss}(2022)}]{stano2022review}%
  \BibitemOpen
  \bibfield  {author} {\bibinfo {author} {\bibfnamefont {P.}~\bibnamefont {Stano}}\ and\ \bibinfo {author} {\bibfnamefont {D.}~\bibnamefont {Loss}},\ }\bibfield  {title} {\bibinfo {title} {Review of performance metrics of spin qubits in gated semiconducting nanostructures},\ }\href@noop {} {\bibfield  {journal} {\bibinfo  {journal} {Nature Reviews Physics}\ }\textbf {\bibinfo {volume} {4}},\ \bibinfo {pages} {672} (\bibinfo {year} {2022})}\BibitemShut {NoStop}%
\bibitem [{\citenamefont {Kuhlmann}\ \emph {et~al.}(2013)\citenamefont {Kuhlmann}, \citenamefont {Houel}, \citenamefont {Ludwig}, \citenamefont {Greuter}, \citenamefont {Reuter}, \citenamefont {Wieck}, \citenamefont {Poggio},\ and\ \citenamefont {Warburton}}]{kuhlmann2013charge}%
  \BibitemOpen
  \bibfield  {author} {\bibinfo {author} {\bibfnamefont {A.~V.}\ \bibnamefont {Kuhlmann}}, \bibinfo {author} {\bibfnamefont {J.}~\bibnamefont {Houel}}, \bibinfo {author} {\bibfnamefont {A.}~\bibnamefont {Ludwig}}, \bibinfo {author} {\bibfnamefont {L.}~\bibnamefont {Greuter}}, \bibinfo {author} {\bibfnamefont {D.}~\bibnamefont {Reuter}}, \bibinfo {author} {\bibfnamefont {A.~D.}\ \bibnamefont {Wieck}}, \bibinfo {author} {\bibfnamefont {M.}~\bibnamefont {Poggio}},\ and\ \bibinfo {author} {\bibfnamefont {R.~J.}\ \bibnamefont {Warburton}},\ }\bibfield  {title} {\bibinfo {title} {Charge noise and spin noise in a semiconductor quantum device},\ }\href@noop {} {\bibfield  {journal} {\bibinfo  {journal} {Nature Physics}\ }\textbf {\bibinfo {volume} {9}},\ \bibinfo {pages} {570} (\bibinfo {year} {2013})}\BibitemShut {NoStop}%
\bibitem [{\citenamefont {Dial}\ \emph {et~al.}(2013)\citenamefont {Dial}, \citenamefont {Shulman}, \citenamefont {Harvey}, \citenamefont {Bluhm}, \citenamefont {Umansky},\ and\ \citenamefont {Yacoby}}]{Dial:2013p146804}%
  \BibitemOpen
  \bibfield  {author} {\bibinfo {author} {\bibfnamefont {O.~E.}\ \bibnamefont {Dial}}, \bibinfo {author} {\bibfnamefont {M.~D.}\ \bibnamefont {Shulman}}, \bibinfo {author} {\bibfnamefont {S.~P.}\ \bibnamefont {Harvey}}, \bibinfo {author} {\bibfnamefont {H.}~\bibnamefont {Bluhm}}, \bibinfo {author} {\bibfnamefont {V.}~\bibnamefont {Umansky}},\ and\ \bibinfo {author} {\bibfnamefont {A.}~\bibnamefont {Yacoby}},\ }\bibfield  {title} {\bibinfo {title} {Charge noise spectroscopy using coherent exchange oscillations in a singlet-triplet qubit},\ }\href@noop {} {\bibfield  {journal} {\bibinfo  {journal} {Phys. Rev. Lett.}\ }\textbf {\bibinfo {volume} {110}},\ \bibinfo {pages} {146804} (\bibinfo {year} {2013})}\BibitemShut {NoStop}%
\bibitem [{\citenamefont {Connors}\ \emph {et~al.}(2022)\citenamefont {Connors}, \citenamefont {Nelson}, \citenamefont {Edge},\ and\ \citenamefont {Nichol}}]{connors2022charge}%
  \BibitemOpen
  \bibfield  {author} {\bibinfo {author} {\bibfnamefont {E.~J.}\ \bibnamefont {Connors}}, \bibinfo {author} {\bibfnamefont {J.}~\bibnamefont {Nelson}}, \bibinfo {author} {\bibfnamefont {L.~F.}\ \bibnamefont {Edge}},\ and\ \bibinfo {author} {\bibfnamefont {J.~M.}\ \bibnamefont {Nichol}},\ }\bibfield  {title} {\bibinfo {title} {Charge-noise spectroscopy of {Si/SiGe} quantum dots via dynamically-decoupled exchange oscillations},\ }\href@noop {} {\bibfield  {journal} {\bibinfo  {journal} {Nature Communications}\ }\textbf {\bibinfo {volume} {13}},\ \bibinfo {pages} {940} (\bibinfo {year} {2022})}\BibitemShut {NoStop}%
\bibitem [{\citenamefont {Yoneda}\ \emph {et~al.}(2018)\citenamefont {Yoneda}, \citenamefont {Takeda}, \citenamefont {Otsuka}, \citenamefont {Nakajima}, \citenamefont {Delbecq}, \citenamefont {Allison}, \citenamefont {Honda}, \citenamefont {Kodera}, \citenamefont {Oda}, \citenamefont {Hoshi} \emph {et~al.}}]{yoneda2018quantum}%
  \BibitemOpen
  \bibfield  {author} {\bibinfo {author} {\bibfnamefont {J.}~\bibnamefont {Yoneda}}, \bibinfo {author} {\bibfnamefont {K.}~\bibnamefont {Takeda}}, \bibinfo {author} {\bibfnamefont {T.}~\bibnamefont {Otsuka}}, \bibinfo {author} {\bibfnamefont {T.}~\bibnamefont {Nakajima}}, \bibinfo {author} {\bibfnamefont {M.~R.}\ \bibnamefont {Delbecq}}, \bibinfo {author} {\bibfnamefont {G.}~\bibnamefont {Allison}}, \bibinfo {author} {\bibfnamefont {T.}~\bibnamefont {Honda}}, \bibinfo {author} {\bibfnamefont {T.}~\bibnamefont {Kodera}}, \bibinfo {author} {\bibfnamefont {S.}~\bibnamefont {Oda}}, \bibinfo {author} {\bibfnamefont {Y.}~\bibnamefont {Hoshi}}, \emph {et~al.},\ }\bibfield  {title} {\bibinfo {title} {A quantum-dot spin qubit with coherence limited by charge noise and fidelity higher than 99.9\%},\ }\href@noop {} {\bibfield  {journal} {\bibinfo  {journal} {Nature Nanotechnology}\ }\textbf {\bibinfo {volume} {13}},\ \bibinfo {pages} {102} (\bibinfo {year} {2018})}\BibitemShut {NoStop}%
\bibitem [{\citenamefont {Struck}\ \emph {et~al.}(2020)\citenamefont {Struck}, \citenamefont {Hollmann}, \citenamefont {Schauer}, \citenamefont {Fedorets}, \citenamefont {Schmidbauer}, \citenamefont {Sawano}, \citenamefont {Riemann}, \citenamefont {Abrosimov}, \citenamefont {Cywi{\'n}ski}, \citenamefont {Bougeard} \emph {et~al.}}]{struck2020low}%
  \BibitemOpen
  \bibfield  {author} {\bibinfo {author} {\bibfnamefont {T.}~\bibnamefont {Struck}}, \bibinfo {author} {\bibfnamefont {A.}~\bibnamefont {Hollmann}}, \bibinfo {author} {\bibfnamefont {F.}~\bibnamefont {Schauer}}, \bibinfo {author} {\bibfnamefont {O.}~\bibnamefont {Fedorets}}, \bibinfo {author} {\bibfnamefont {A.}~\bibnamefont {Schmidbauer}}, \bibinfo {author} {\bibfnamefont {K.}~\bibnamefont {Sawano}}, \bibinfo {author} {\bibfnamefont {H.}~\bibnamefont {Riemann}}, \bibinfo {author} {\bibfnamefont {N.~V.}\ \bibnamefont {Abrosimov}}, \bibinfo {author} {\bibfnamefont {{\L}.}~\bibnamefont {Cywi{\'n}ski}}, \bibinfo {author} {\bibfnamefont {D.}~\bibnamefont {Bougeard}}, \emph {et~al.},\ }\bibfield  {title} {\bibinfo {title} {Low-frequency spin qubit energy splitting noise in highly purified $^{28}${Si/SiGe}},\ }\href@noop {} {\bibfield  {journal} {\bibinfo  {journal} {npj Quantum Information}\ }\textbf {\bibinfo {volume} {6}},\ \bibinfo {pages} {40} (\bibinfo {year} {2020})}\BibitemShut {NoStop}%
\bibitem [{\citenamefont {Paladino}\ \emph {et~al.}(2014)\citenamefont {Paladino}, \citenamefont {Galperin}, \citenamefont {Falci},\ and\ \citenamefont {Altshuler}}]{paladino20141}%
  \BibitemOpen
  \bibfield  {author} {\bibinfo {author} {\bibfnamefont {E.}~\bibnamefont {Paladino}}, \bibinfo {author} {\bibfnamefont {Y.}~\bibnamefont {Galperin}}, \bibinfo {author} {\bibfnamefont {G.}~\bibnamefont {Falci}},\ and\ \bibinfo {author} {\bibfnamefont {B.}~\bibnamefont {Altshuler}},\ }\bibfield  {title} {\bibinfo {title} {1/f noise: Implications for solid-state quantum information},\ }\href@noop {} {\bibfield  {journal} {\bibinfo  {journal} {Reviews of Modern Physics}\ }\textbf {\bibinfo {volume} {86}},\ \bibinfo {pages} {361} (\bibinfo {year} {2014})}\BibitemShut {NoStop}%
\bibitem [{\citenamefont {Connors}\ \emph {et~al.}(2019)\citenamefont {Connors}, \citenamefont {Nelson}, \citenamefont {Qiao}, \citenamefont {Edge},\ and\ \citenamefont {Nichol}}]{connors2019low}%
  \BibitemOpen
  \bibfield  {author} {\bibinfo {author} {\bibfnamefont {E.~J.}\ \bibnamefont {Connors}}, \bibinfo {author} {\bibfnamefont {J.}~\bibnamefont {Nelson}}, \bibinfo {author} {\bibfnamefont {H.}~\bibnamefont {Qiao}}, \bibinfo {author} {\bibfnamefont {L.~F.}\ \bibnamefont {Edge}},\ and\ \bibinfo {author} {\bibfnamefont {J.~M.}\ \bibnamefont {Nichol}},\ }\bibfield  {title} {\bibinfo {title} {Low-frequency charge noise in {S}i/{S}i{G}e quantum dots},\ }\href@noop {} {\bibfield  {journal} {\bibinfo  {journal} {Phys. Rev. B}\ }\textbf {\bibinfo {volume} {100}},\ \bibinfo {pages} {165305} (\bibinfo {year} {2019})}\BibitemShut {NoStop}%
\bibitem [{\citenamefont {M{\"u}ller}\ \emph {et~al.}(2019)\citenamefont {M{\"u}ller}, \citenamefont {Cole},\ and\ \citenamefont {Lisenfeld}}]{muller2019towards}%
  \BibitemOpen
  \bibfield  {author} {\bibinfo {author} {\bibfnamefont {C.}~\bibnamefont {M{\"u}ller}}, \bibinfo {author} {\bibfnamefont {J.~H.}\ \bibnamefont {Cole}},\ and\ \bibinfo {author} {\bibfnamefont {J.}~\bibnamefont {Lisenfeld}},\ }\bibfield  {title} {\bibinfo {title} {Towards understanding two-level-systems in amorphous solids: insights from quantum circuits},\ }\href@noop {} {\bibfield  {journal} {\bibinfo  {journal} {Reports on Progress in Physics}\ }\textbf {\bibinfo {volume} {82}},\ \bibinfo {pages} {124501} (\bibinfo {year} {2019})}\BibitemShut {NoStop}%
\bibitem [{\citenamefont {Ye}\ \emph {et~al.}(2024)\citenamefont {Ye}, \citenamefont {Ellaboudy}, \citenamefont {Albrecht}, \citenamefont {Vudatha}, \citenamefont {Jacobson},\ and\ \citenamefont {Nichol}}]{ye2024characterization}%
  \BibitemOpen
  \bibfield  {author} {\bibinfo {author} {\bibfnamefont {F.}~\bibnamefont {Ye}}, \bibinfo {author} {\bibfnamefont {A.}~\bibnamefont {Ellaboudy}}, \bibinfo {author} {\bibfnamefont {D.}~\bibnamefont {Albrecht}}, \bibinfo {author} {\bibfnamefont {R.}~\bibnamefont {Vudatha}}, \bibinfo {author} {\bibfnamefont {N.~T.}\ \bibnamefont {Jacobson}},\ and\ \bibinfo {author} {\bibfnamefont {J.~M.}\ \bibnamefont {Nichol}},\ }\bibfield  {title} {\bibinfo {title} {Characterization of individual charge fluctuators in si/sige quantum dots},\ }\href@noop {} {\bibfield  {journal} {\bibinfo  {journal} {arXiv preprint arXiv:2401.14541}\ } (\bibinfo {year} {2024})}\BibitemShut {NoStop}%
\bibitem [{\citenamefont {Devitt}\ \emph {et~al.}(2013)\citenamefont {Devitt}, \citenamefont {Munro},\ and\ \citenamefont {Nemoto}}]{devitt2013quantum}%
  \BibitemOpen
  \bibfield  {author} {\bibinfo {author} {\bibfnamefont {S.~J.}\ \bibnamefont {Devitt}}, \bibinfo {author} {\bibfnamefont {W.~J.}\ \bibnamefont {Munro}},\ and\ \bibinfo {author} {\bibfnamefont {K.}~\bibnamefont {Nemoto}},\ }\bibfield  {title} {\bibinfo {title} {Quantum error correction for beginners},\ }\href@noop {} {\bibfield  {journal} {\bibinfo  {journal} {Reports on Progress in Physics}\ }\textbf {\bibinfo {volume} {76}},\ \bibinfo {pages} {076001} (\bibinfo {year} {2013})}\BibitemShut {NoStop}%
\bibitem [{\citenamefont {Roffe}(2019)}]{roffe2019quantum}%
  \BibitemOpen
  \bibfield  {author} {\bibinfo {author} {\bibfnamefont {J.}~\bibnamefont {Roffe}},\ }\bibfield  {title} {\bibinfo {title} {Quantum error correction: an introductory guide},\ }\href@noop {} {\bibfield  {journal} {\bibinfo  {journal} {Contemporary Physics}\ }\textbf {\bibinfo {volume} {60}},\ \bibinfo {pages} {226} (\bibinfo {year} {2019})}\BibitemShut {NoStop}%
\bibitem [{\citenamefont {Endo}\ \emph {et~al.}(2018)\citenamefont {Endo}, \citenamefont {Benjamin},\ and\ \citenamefont {Li}}]{endo2018practical}%
  \BibitemOpen
  \bibfield  {author} {\bibinfo {author} {\bibfnamefont {S.}~\bibnamefont {Endo}}, \bibinfo {author} {\bibfnamefont {S.~C.}\ \bibnamefont {Benjamin}},\ and\ \bibinfo {author} {\bibfnamefont {Y.}~\bibnamefont {Li}},\ }\bibfield  {title} {\bibinfo {title} {Practical quantum error mitigation for near-future applications},\ }\href@noop {} {\bibfield  {journal} {\bibinfo  {journal} {Physical Review X}\ }\textbf {\bibinfo {volume} {8}},\ \bibinfo {pages} {031027} (\bibinfo {year} {2018})}\BibitemShut {NoStop}%
\bibitem [{\citenamefont {Strikis}\ \emph {et~al.}(2021)\citenamefont {Strikis}, \citenamefont {Qin}, \citenamefont {Chen}, \citenamefont {Benjamin},\ and\ \citenamefont {Li}}]{StrikisPRXQuantum2021}%
  \BibitemOpen
  \bibfield  {author} {\bibinfo {author} {\bibfnamefont {A.}~\bibnamefont {Strikis}}, \bibinfo {author} {\bibfnamefont {D.}~\bibnamefont {Qin}}, \bibinfo {author} {\bibfnamefont {Y.}~\bibnamefont {Chen}}, \bibinfo {author} {\bibfnamefont {S.~C.}\ \bibnamefont {Benjamin}},\ and\ \bibinfo {author} {\bibfnamefont {Y.}~\bibnamefont {Li}},\ }\bibfield  {title} {\bibinfo {title} {Learning-based quantum error mitigation},\ }\href@noop {} {\bibfield  {journal} {\bibinfo  {journal} {PRX Quantum}\ }\textbf {\bibinfo {volume} {2}},\ \bibinfo {pages} {040330} (\bibinfo {year} {2021})}\BibitemShut {NoStop}%
\bibitem [{\citenamefont {Suter}\ and\ \citenamefont {{\'A}lvarez}(2016)}]{suter2016colloquium}%
  \BibitemOpen
  \bibfield  {author} {\bibinfo {author} {\bibfnamefont {D.}~\bibnamefont {Suter}}\ and\ \bibinfo {author} {\bibfnamefont {G.~A.}\ \bibnamefont {{\'A}lvarez}},\ }\bibfield  {title} {\bibinfo {title} {Colloquium: Protecting quantum information against environmental noise},\ }\href@noop {} {\bibfield  {journal} {\bibinfo  {journal} {Reviews of Modern Physics}\ }\textbf {\bibinfo {volume} {88}},\ \bibinfo {pages} {041001} (\bibinfo {year} {2016})}\BibitemShut {NoStop}%
\bibitem [{\citenamefont {Lidar}(2014)}]{lidar2014review}%
  \BibitemOpen
  \bibfield  {author} {\bibinfo {author} {\bibfnamefont {D.~A.}\ \bibnamefont {Lidar}},\ }\bibfield  {title} {\bibinfo {title} {Review of decoherence-free subspaces, noiseless subsystems, and dynamical decoupling},\ }\href@noop {} {\bibfield  {journal} {\bibinfo  {journal} {Quantum Information and Computation for Chemistry}\ ,\ \bibinfo {pages} {295}} (\bibinfo {year} {2014})}\BibitemShut {NoStop}%
\bibitem [{\citenamefont {Singh}\ \emph {et~al.}(2023)\citenamefont {Singh}, \citenamefont {Bradley}, \citenamefont {Anand}, \citenamefont {Ramesh}, \citenamefont {White},\ and\ \citenamefont {Bernien}}]{singh2023mid}%
  \BibitemOpen
  \bibfield  {author} {\bibinfo {author} {\bibfnamefont {K.}~\bibnamefont {Singh}}, \bibinfo {author} {\bibfnamefont {C.~E.}\ \bibnamefont {Bradley}}, \bibinfo {author} {\bibfnamefont {S.}~\bibnamefont {Anand}}, \bibinfo {author} {\bibfnamefont {V.}~\bibnamefont {Ramesh}}, \bibinfo {author} {\bibfnamefont {R.}~\bibnamefont {White}},\ and\ \bibinfo {author} {\bibfnamefont {H.}~\bibnamefont {Bernien}},\ }\bibfield  {title} {\bibinfo {title} {Mid-circuit correction of correlated phase errors using an array of spectator qubits},\ }\href@noop {} {\bibfield  {journal} {\bibinfo  {journal} {Science}\ }\textbf {\bibinfo {volume} {380}},\ \bibinfo {pages} {1265} (\bibinfo {year} {2023})}\BibitemShut {NoStop}%
\bibitem [{\citenamefont {Barnes}\ \emph {et~al.}(2022)\citenamefont {Barnes}, \citenamefont {Calderon-Vargas}, \citenamefont {Dong}, \citenamefont {Li}, \citenamefont {Zeng},\ and\ \citenamefont {Zhuang}}]{barnes2022dynamically}%
  \BibitemOpen
  \bibfield  {author} {\bibinfo {author} {\bibfnamefont {E.}~\bibnamefont {Barnes}}, \bibinfo {author} {\bibfnamefont {F.~A.}\ \bibnamefont {Calderon-Vargas}}, \bibinfo {author} {\bibfnamefont {W.}~\bibnamefont {Dong}}, \bibinfo {author} {\bibfnamefont {B.}~\bibnamefont {Li}}, \bibinfo {author} {\bibfnamefont {J.}~\bibnamefont {Zeng}},\ and\ \bibinfo {author} {\bibfnamefont {F.}~\bibnamefont {Zhuang}},\ }\bibfield  {title} {\bibinfo {title} {Dynamically corrected gates from geometric space curves},\ }\href@noop {} {\bibfield  {journal} {\bibinfo  {journal} {Quantum Science and Technology}\ }\textbf {\bibinfo {volume} {7}},\ \bibinfo {pages} {023001} (\bibinfo {year} {2022})}\BibitemShut {NoStop}%
\bibitem [{\citenamefont {David}(2015)}]{david2015vessels}%
  \BibitemOpen
  \bibfield  {author} {\bibinfo {author} {\bibfnamefont {M.}~\bibnamefont {David}},\ }\bibfield  {title} {\bibinfo {title} {Vessels and ballast water},\ }\href@noop {} {\bibfield  {journal} {\bibinfo  {journal} {Global Maritime Transport and Ballast Water Management: Issues and Solutions}\ ,\ \bibinfo {pages} {13}} (\bibinfo {year} {2015})}\BibitemShut {NoStop}%
\bibitem [{\citenamefont {Paquelet~Wuetz}\ \emph {et~al.}(2023)\citenamefont {Paquelet~Wuetz}, \citenamefont {Degli~Esposti}, \citenamefont {Zwerver}, \citenamefont {Amitonov}, \citenamefont {Botifoll}, \citenamefont {Arbiol}, \citenamefont {Sammak}, \citenamefont {Vandersypen}, \citenamefont {Russ},\ and\ \citenamefont {Scappucci}}]{paquelet2023reducing}%
  \BibitemOpen
  \bibfield  {author} {\bibinfo {author} {\bibfnamefont {B.}~\bibnamefont {Paquelet~Wuetz}}, \bibinfo {author} {\bibfnamefont {D.}~\bibnamefont {Degli~Esposti}}, \bibinfo {author} {\bibfnamefont {A.-M.~J.}\ \bibnamefont {Zwerver}}, \bibinfo {author} {\bibfnamefont {S.~V.}\ \bibnamefont {Amitonov}}, \bibinfo {author} {\bibfnamefont {M.}~\bibnamefont {Botifoll}}, \bibinfo {author} {\bibfnamefont {J.}~\bibnamefont {Arbiol}}, \bibinfo {author} {\bibfnamefont {A.}~\bibnamefont {Sammak}}, \bibinfo {author} {\bibfnamefont {L.~M.}\ \bibnamefont {Vandersypen}}, \bibinfo {author} {\bibfnamefont {M.}~\bibnamefont {Russ}},\ and\ \bibinfo {author} {\bibfnamefont {G.}~\bibnamefont {Scappucci}},\ }\bibfield  {title} {\bibinfo {title} {Reducing charge noise in quantum dots by using thin silicon quantum wells},\ }\href@noop {} {\bibfield  {journal} {\bibinfo  {journal} {Nature communications}\ }\textbf {\bibinfo {volume} {14}},\ \bibinfo {pages} {1385} (\bibinfo {year} {2023})}\BibitemShut {NoStop}%
\bibitem [{\citenamefont {Barnes}\ \emph {et~al.}(2011)\citenamefont {Barnes}, \citenamefont {Kestner}, \citenamefont {Nguyen},\ and\ \citenamefont {Sarma}}]{barnes2011screening}%
  \BibitemOpen
  \bibfield  {author} {\bibinfo {author} {\bibfnamefont {E.}~\bibnamefont {Barnes}}, \bibinfo {author} {\bibfnamefont {J.}~\bibnamefont {Kestner}}, \bibinfo {author} {\bibfnamefont {N.}~\bibnamefont {Nguyen}},\ and\ \bibinfo {author} {\bibfnamefont {S.~D.}\ \bibnamefont {Sarma}},\ }\bibfield  {title} {\bibinfo {title} {Screening of charged impurities with multielectron singlet-triplet spin qubits in quantum dots},\ }\href@noop {} {\bibfield  {journal} {\bibinfo  {journal} {Physical Review B}\ }\textbf {\bibinfo {volume} {84}},\ \bibinfo {pages} {235309} (\bibinfo {year} {2011})}\BibitemShut {NoStop}%
\bibitem [{\citenamefont {Higginbotham}\ \emph {et~al.}(2014)\citenamefont {Higginbotham}, \citenamefont {Kuemmeth}, \citenamefont {Hanson}, \citenamefont {Gossard},\ and\ \citenamefont {Marcus}}]{higginbotham2014coherent}%
  \BibitemOpen
  \bibfield  {author} {\bibinfo {author} {\bibfnamefont {A.~P.}\ \bibnamefont {Higginbotham}}, \bibinfo {author} {\bibfnamefont {F.}~\bibnamefont {Kuemmeth}}, \bibinfo {author} {\bibfnamefont {M.~P.}\ \bibnamefont {Hanson}}, \bibinfo {author} {\bibfnamefont {A.~C.}\ \bibnamefont {Gossard}},\ and\ \bibinfo {author} {\bibfnamefont {C.~M.}\ \bibnamefont {Marcus}},\ }\bibfield  {title} {\bibinfo {title} {Coherent operations and screening in multielectron spin qubits},\ }\href@noop {} {\bibfield  {journal} {\bibinfo  {journal} {Physical Review Letters}\ }\textbf {\bibinfo {volume} {112}},\ \bibinfo {pages} {026801} (\bibinfo {year} {2014})}\BibitemShut {NoStop}%
\bibitem [{\citenamefont {Schriefl}\ \emph {et~al.}(2006)\citenamefont {Schriefl}, \citenamefont {Makhlin}, \citenamefont {Shnirman},\ and\ \citenamefont {Sch{\"o}n}}]{schriefl2006decoherence}%
  \BibitemOpen
  \bibfield  {author} {\bibinfo {author} {\bibfnamefont {J.}~\bibnamefont {Schriefl}}, \bibinfo {author} {\bibfnamefont {Y.}~\bibnamefont {Makhlin}}, \bibinfo {author} {\bibfnamefont {A.}~\bibnamefont {Shnirman}},\ and\ \bibinfo {author} {\bibfnamefont {G.}~\bibnamefont {Sch{\"o}n}},\ }\bibfield  {title} {\bibinfo {title} {Decoherence from ensembles of two-level fluctuators},\ }\href@noop {} {\bibfield  {journal} {\bibinfo  {journal} {New Journal of Physics}\ }\textbf {\bibinfo {volume} {8}},\ \bibinfo {pages} {1} (\bibinfo {year} {2006})}\BibitemShut {NoStop}%
\bibitem [{\citenamefont {Choi}\ and\ \citenamefont {Joynt}(2022)}]{choi2022anisotropy}%
  \BibitemOpen
  \bibfield  {author} {\bibinfo {author} {\bibfnamefont {Y.}~\bibnamefont {Choi}}\ and\ \bibinfo {author} {\bibfnamefont {R.}~\bibnamefont {Joynt}},\ }\bibfield  {title} {\bibinfo {title} {Anisotropy with respect to the applied magnetic field of spin qubit decoherence times},\ }\href@noop {} {\bibfield  {journal} {\bibinfo  {journal} {npj Quantum Information}\ }\textbf {\bibinfo {volume} {8}},\ \bibinfo {pages} {70} (\bibinfo {year} {2022})}\BibitemShut {NoStop}%
\bibitem [{\citenamefont {Ashlea~Alava}\ \emph {et~al.}(2021)\citenamefont {Ashlea~Alava}, \citenamefont {Wang}, \citenamefont {Chen}, \citenamefont {Ritchie}, \citenamefont {Klochan},\ and\ \citenamefont {Hamilton}}]{ashlea2021high}%
  \BibitemOpen
  \bibfield  {author} {\bibinfo {author} {\bibfnamefont {Y.}~\bibnamefont {Ashlea~Alava}}, \bibinfo {author} {\bibfnamefont {D.}~\bibnamefont {Wang}}, \bibinfo {author} {\bibfnamefont {C.}~\bibnamefont {Chen}}, \bibinfo {author} {\bibfnamefont {D.}~\bibnamefont {Ritchie}}, \bibinfo {author} {\bibfnamefont {O.}~\bibnamefont {Klochan}},\ and\ \bibinfo {author} {\bibfnamefont {A.}~\bibnamefont {Hamilton}},\ }\bibfield  {title} {\bibinfo {title} {High electron mobility and low noise quantum point contacts in an ultra-shallow all-epitaxial metal gate gaas/alxga1- xas heterostructure},\ }\href@noop {} {\bibfield  {journal} {\bibinfo  {journal} {Applied Physics Letters}\ }\textbf {\bibinfo {volume} {119}} (\bibinfo {year} {2021})}\BibitemShut {NoStop}%
\bibitem [{\citenamefont {Cheng}\ \emph {et~al.}(2016)\citenamefont {Cheng}, \citenamefont {Su}, \citenamefont {Choi}, \citenamefont {Gwo}, \citenamefont {Li},\ and\ \citenamefont {Shih}}]{cheng2016epitaxial}%
  \BibitemOpen
  \bibfield  {author} {\bibinfo {author} {\bibfnamefont {F.}~\bibnamefont {Cheng}}, \bibinfo {author} {\bibfnamefont {P.-H.}\ \bibnamefont {Su}}, \bibinfo {author} {\bibfnamefont {J.}~\bibnamefont {Choi}}, \bibinfo {author} {\bibfnamefont {S.}~\bibnamefont {Gwo}}, \bibinfo {author} {\bibfnamefont {X.}~\bibnamefont {Li}},\ and\ \bibinfo {author} {\bibfnamefont {C.-K.}\ \bibnamefont {Shih}},\ }\bibfield  {title} {\bibinfo {title} {Epitaxial growth of atomically smooth aluminum on silicon and its intrinsic optical properties},\ }\href@noop {} {\bibfield  {journal} {\bibinfo  {journal} {ACS nano}\ }\textbf {\bibinfo {volume} {10}},\ \bibinfo {pages} {9852} (\bibinfo {year} {2016})}\BibitemShut {NoStop}%
\bibitem [{\citenamefont {Nakamura}\ \emph {et~al.}(2019)\citenamefont {Nakamura}, \citenamefont {Fallahi}, \citenamefont {Sahasrabudhe}, \citenamefont {Rahman}, \citenamefont {Liang}, \citenamefont {Gardner},\ and\ \citenamefont {Manfra}}]{nakamura2019aharonov}%
  \BibitemOpen
  \bibfield  {author} {\bibinfo {author} {\bibfnamefont {J.}~\bibnamefont {Nakamura}}, \bibinfo {author} {\bibfnamefont {S.}~\bibnamefont {Fallahi}}, \bibinfo {author} {\bibfnamefont {H.}~\bibnamefont {Sahasrabudhe}}, \bibinfo {author} {\bibfnamefont {R.}~\bibnamefont {Rahman}}, \bibinfo {author} {\bibfnamefont {S.}~\bibnamefont {Liang}}, \bibinfo {author} {\bibfnamefont {G.~C.}\ \bibnamefont {Gardner}},\ and\ \bibinfo {author} {\bibfnamefont {M.~J.}\ \bibnamefont {Manfra}},\ }\bibfield  {title} {\bibinfo {title} {Aharonov--bohm interference of fractional quantum hall edge modes},\ }\href@noop {} {\bibfield  {journal} {\bibinfo  {journal} {Nature Physics}\ }\textbf {\bibinfo {volume} {15}},\ \bibinfo {pages} {563} (\bibinfo {year} {2019})}\BibitemShut {NoStop}%
\bibitem [{\citenamefont {Borselli}\ \emph {et~al.}(2011)\citenamefont {Borselli}, \citenamefont {Ross}, \citenamefont {Kiselev}, \citenamefont {Croke}, \citenamefont {Holabird}, \citenamefont {Deelman}, \citenamefont {Warren}, \citenamefont {Alvarado-Rodriguez}, \citenamefont {Milosavljevic}, \citenamefont {Ku}, \citenamefont {Wong}, \citenamefont {Schmitz}, \citenamefont {Sokolich}, \citenamefont {Gyure},\ and\ \citenamefont {Hunter}}]{Borselli:2011p123118}%
  \BibitemOpen
  \bibfield  {author} {\bibinfo {author} {\bibfnamefont {M.~G.}\ \bibnamefont {Borselli}}, \bibinfo {author} {\bibfnamefont {R.~S.}\ \bibnamefont {Ross}}, \bibinfo {author} {\bibfnamefont {A.~A.}\ \bibnamefont {Kiselev}}, \bibinfo {author} {\bibfnamefont {E.~T.}\ \bibnamefont {Croke}}, \bibinfo {author} {\bibfnamefont {K.~S.}\ \bibnamefont {Holabird}}, \bibinfo {author} {\bibfnamefont {P.~W.}\ \bibnamefont {Deelman}}, \bibinfo {author} {\bibfnamefont {L.~D.}\ \bibnamefont {Warren}}, \bibinfo {author} {\bibfnamefont {I.}~\bibnamefont {Alvarado-Rodriguez}}, \bibinfo {author} {\bibfnamefont {I.}~\bibnamefont {Milosavljevic}}, \bibinfo {author} {\bibfnamefont {F.~C.}\ \bibnamefont {Ku}}, \bibinfo {author} {\bibfnamefont {W.~S.}\ \bibnamefont {Wong}}, \bibinfo {author} {\bibfnamefont {A.~E.}\ \bibnamefont {Schmitz}}, \bibinfo {author} {\bibfnamefont {M.}~\bibnamefont {Sokolich}}, \bibinfo {author} {\bibfnamefont {M.~F.}\ \bibnamefont {Gyure}},\ and\ \bibinfo {author} {\bibfnamefont {A.~T.}\ \bibnamefont
  {Hunter}},\ }\bibfield  {title} {\bibinfo {title} {Measurement of valley splitting in high-symmetry {S}i/{S}i{G}e quantum dots},\ }\href@noop {} {\bibfield  {journal} {\bibinfo  {journal} {Appl. Phys. Lett.}\ }\textbf {\bibinfo {volume} {98}},\ \bibinfo {pages} {123118} (\bibinfo {year} {2011})}\BibitemShut {NoStop}%
\bibitem [{\citenamefont {Ivlev}\ \emph {et~al.}(2024)\citenamefont {Ivlev}, \citenamefont {Tidjani}, \citenamefont {Oosterhout}, \citenamefont {Sammak}, \citenamefont {Scappucci},\ and\ \citenamefont {Veldhorst}}]{ivlev2024coupled}%
  \BibitemOpen
  \bibfield  {author} {\bibinfo {author} {\bibfnamefont {A.}~\bibnamefont {Ivlev}}, \bibinfo {author} {\bibfnamefont {H.}~\bibnamefont {Tidjani}}, \bibinfo {author} {\bibfnamefont {S.}~\bibnamefont {Oosterhout}}, \bibinfo {author} {\bibfnamefont {A.}~\bibnamefont {Sammak}}, \bibinfo {author} {\bibfnamefont {G.}~\bibnamefont {Scappucci}},\ and\ \bibinfo {author} {\bibfnamefont {M.}~\bibnamefont {Veldhorst}},\ }\bibfield  {title} {\bibinfo {title} {Coupled vertical double quantum dots at single-hole occupancy},\ }\href@noop {} {\bibfield  {journal} {\bibinfo  {journal} {arXiv preprint arXiv:2401.07736}\ } (\bibinfo {year} {2024})}\BibitemShut {NoStop}%
\bibitem [{\citenamefont {McJunkin}\ \emph {et~al.}(2021)\citenamefont {McJunkin}, \citenamefont {MacQuarrie}, \citenamefont {Tom}, \citenamefont {Neyens}, \citenamefont {Dodson}, \citenamefont {Thorgrimsson}, \citenamefont {Corrigan}, \citenamefont {Ercan}, \citenamefont {Savage}, \citenamefont {Lagally} \emph {et~al.}}]{mcjunkin2021valley}%
  \BibitemOpen
  \bibfield  {author} {\bibinfo {author} {\bibfnamefont {T.}~\bibnamefont {McJunkin}}, \bibinfo {author} {\bibfnamefont {E.}~\bibnamefont {MacQuarrie}}, \bibinfo {author} {\bibfnamefont {L.}~\bibnamefont {Tom}}, \bibinfo {author} {\bibfnamefont {S.}~\bibnamefont {Neyens}}, \bibinfo {author} {\bibfnamefont {J.}~\bibnamefont {Dodson}}, \bibinfo {author} {\bibfnamefont {B.}~\bibnamefont {Thorgrimsson}}, \bibinfo {author} {\bibfnamefont {J.}~\bibnamefont {Corrigan}}, \bibinfo {author} {\bibfnamefont {H.~E.}\ \bibnamefont {Ercan}}, \bibinfo {author} {\bibfnamefont {D.}~\bibnamefont {Savage}}, \bibinfo {author} {\bibfnamefont {M.}~\bibnamefont {Lagally}}, \emph {et~al.},\ }\bibfield  {title} {\bibinfo {title} {Valley splittings in {S}i/{S}i{G}e quantum dots with a germanium spike in the silicon well},\ }\href@noop {} {\bibfield  {journal} {\bibinfo  {journal} {Physical Review B}\ }\textbf {\bibinfo {volume} {104}},\ \bibinfo {pages} {085406} (\bibinfo {year} {2021})}\BibitemShut {NoStop}%
\bibitem [{\citenamefont {Vandersypen}\ \emph {et~al.}(2017)\citenamefont {Vandersypen}, \citenamefont {Bluhm}, \citenamefont {Clarke}, \citenamefont {Dzurak}, \citenamefont {Ishihara}, \citenamefont {Morello}, \citenamefont {Reilly}, \citenamefont {Schreiber},\ and\ \citenamefont {Veldhorst}}]{vandersypen2017interfacing}%
  \BibitemOpen
  \bibfield  {author} {\bibinfo {author} {\bibfnamefont {L.}~\bibnamefont {Vandersypen}}, \bibinfo {author} {\bibfnamefont {H.}~\bibnamefont {Bluhm}}, \bibinfo {author} {\bibfnamefont {J.}~\bibnamefont {Clarke}}, \bibinfo {author} {\bibfnamefont {A.}~\bibnamefont {Dzurak}}, \bibinfo {author} {\bibfnamefont {R.}~\bibnamefont {Ishihara}}, \bibinfo {author} {\bibfnamefont {A.}~\bibnamefont {Morello}}, \bibinfo {author} {\bibfnamefont {D.}~\bibnamefont {Reilly}}, \bibinfo {author} {\bibfnamefont {L.}~\bibnamefont {Schreiber}},\ and\ \bibinfo {author} {\bibfnamefont {M.}~\bibnamefont {Veldhorst}},\ }\bibfield  {title} {\bibinfo {title} {Interfacing spin qubits in quantum dots and donors---hot, dense, and coherent},\ }\href@noop {} {\bibfield  {journal} {\bibinfo  {journal} {npj Quantum Information}\ }\textbf {\bibinfo {volume} {3}},\ \bibinfo {pages} {34} (\bibinfo {year} {2017})}\BibitemShut {NoStop}%
\bibitem [{\citenamefont {Li}\ \emph {et~al.}(2018)\citenamefont {Li}, \citenamefont {Petit}, \citenamefont {Franke}, \citenamefont {Dehollain}, \citenamefont {Helsen}, \citenamefont {Steudtner}, \citenamefont {Thomas}, \citenamefont {Yoscovits}, \citenamefont {Singh}, \citenamefont {Wehner}, \citenamefont {Vandersypen}, \citenamefont {Clarke},\ and\ \citenamefont {Veldhorst}}]{Li:2018p7}%
  \BibitemOpen
  \bibfield  {author} {\bibinfo {author} {\bibfnamefont {R.}~\bibnamefont {Li}}, \bibinfo {author} {\bibfnamefont {L.}~\bibnamefont {Petit}}, \bibinfo {author} {\bibfnamefont {D.~P.}\ \bibnamefont {Franke}}, \bibinfo {author} {\bibfnamefont {J.~P.}\ \bibnamefont {Dehollain}}, \bibinfo {author} {\bibfnamefont {J.}~\bibnamefont {Helsen}}, \bibinfo {author} {\bibfnamefont {M.}~\bibnamefont {Steudtner}}, \bibinfo {author} {\bibfnamefont {N.~K.}\ \bibnamefont {Thomas}}, \bibinfo {author} {\bibfnamefont {Z.~R.}\ \bibnamefont {Yoscovits}}, \bibinfo {author} {\bibfnamefont {K.~J.}\ \bibnamefont {Singh}}, \bibinfo {author} {\bibfnamefont {S.}~\bibnamefont {Wehner}}, \bibinfo {author} {\bibfnamefont {L.~M.~K.}\ \bibnamefont {Vandersypen}}, \bibinfo {author} {\bibfnamefont {J.~S.}\ \bibnamefont {Clarke}},\ and\ \bibinfo {author} {\bibfnamefont {M.}~\bibnamefont {Veldhorst}},\ }\bibfield  {title} {\bibinfo {title} {A crossbar network for silicon quantum dot qubits},\ }\href@noop {} {\bibfield  {journal} {\bibinfo
  {journal} {Science Advances}\ }\textbf {\bibinfo {volume} {4}},\ \bibinfo {pages} {7} (\bibinfo {year} {2018})}\BibitemShut {NoStop}%
\bibitem [{\citenamefont {Rojas-Arias}\ \emph {et~al.}(2023)\citenamefont {Rojas-Arias}, \citenamefont {Noiri}, \citenamefont {Stano}, \citenamefont {Nakajima}, \citenamefont {Yoneda}, \citenamefont {Takeda}, \citenamefont {Kobayashi}, \citenamefont {Sammak}, \citenamefont {Scappucci}, \citenamefont {Loss} \emph {et~al.}}]{rojas2023spatial}%
  \BibitemOpen
  \bibfield  {author} {\bibinfo {author} {\bibfnamefont {J.~S.}\ \bibnamefont {Rojas-Arias}}, \bibinfo {author} {\bibfnamefont {A.}~\bibnamefont {Noiri}}, \bibinfo {author} {\bibfnamefont {P.}~\bibnamefont {Stano}}, \bibinfo {author} {\bibfnamefont {T.}~\bibnamefont {Nakajima}}, \bibinfo {author} {\bibfnamefont {J.}~\bibnamefont {Yoneda}}, \bibinfo {author} {\bibfnamefont {K.}~\bibnamefont {Takeda}}, \bibinfo {author} {\bibfnamefont {T.}~\bibnamefont {Kobayashi}}, \bibinfo {author} {\bibfnamefont {A.}~\bibnamefont {Sammak}}, \bibinfo {author} {\bibfnamefont {G.}~\bibnamefont {Scappucci}}, \bibinfo {author} {\bibfnamefont {D.}~\bibnamefont {Loss}}, \emph {et~al.},\ }\bibfield  {title} {\bibinfo {title} {Spatial noise correlations beyond nearest neighbors in $^{28}${Si/SiGe} spin qubits},\ }\href@noop {} {\bibfield  {journal} {\bibinfo  {journal} {Physical Review Applied}\ }\textbf {\bibinfo {volume} {20}},\ \bibinfo {pages} {054024} (\bibinfo {year} {2023})}\BibitemShut {NoStop}%
\bibitem [{\citenamefont {Ha}\ \emph {et~al.}(2021)\citenamefont {Ha}, \citenamefont {Ha}, \citenamefont {Choi}, \citenamefont {Tang}, \citenamefont {Schmitz}, \citenamefont {Levendorf}, \citenamefont {Lee}, \citenamefont {Chappell}, \citenamefont {Adams}, \citenamefont {Hulbert} \emph {et~al.}}]{ha2021flexible}%
  \BibitemOpen
  \bibfield  {author} {\bibinfo {author} {\bibfnamefont {W.}~\bibnamefont {Ha}}, \bibinfo {author} {\bibfnamefont {S.~D.}\ \bibnamefont {Ha}}, \bibinfo {author} {\bibfnamefont {M.~D.}\ \bibnamefont {Choi}}, \bibinfo {author} {\bibfnamefont {Y.}~\bibnamefont {Tang}}, \bibinfo {author} {\bibfnamefont {A.~E.}\ \bibnamefont {Schmitz}}, \bibinfo {author} {\bibfnamefont {M.~P.}\ \bibnamefont {Levendorf}}, \bibinfo {author} {\bibfnamefont {K.}~\bibnamefont {Lee}}, \bibinfo {author} {\bibfnamefont {J.~M.}\ \bibnamefont {Chappell}}, \bibinfo {author} {\bibfnamefont {T.~S.}\ \bibnamefont {Adams}}, \bibinfo {author} {\bibfnamefont {D.~R.}\ \bibnamefont {Hulbert}}, \emph {et~al.},\ }\bibfield  {title} {\bibinfo {title} {A flexible design platform for {Si/SiGe} exchange-only qubits with low disorder},\ }\href@noop {} {\bibfield  {journal} {\bibinfo  {journal} {Nano Letters}\ }\textbf {\bibinfo {volume} {22}},\ \bibinfo {pages} {1443} (\bibinfo {year} {2021})}\BibitemShut {NoStop}%
\bibitem [{\citenamefont {Langsjoen}\ \emph {et~al.}(2014)\citenamefont {Langsjoen}, \citenamefont {Poudel}, \citenamefont {Vavilov},\ and\ \citenamefont {Joynt}}]{langsjoen2014}%
  \BibitemOpen
  \bibfield  {author} {\bibinfo {author} {\bibfnamefont {L.}~\bibnamefont {Langsjoen}}, \bibinfo {author} {\bibfnamefont {A.}~\bibnamefont {Poudel}}, \bibinfo {author} {\bibfnamefont {M.}~\bibnamefont {Vavilov}},\ and\ \bibinfo {author} {\bibfnamefont {R.}~\bibnamefont {Joynt}},\ }\bibfield  {title} {\bibinfo {title} {Electromagnetic fluctuations near metallic thin films},\ }\href@noop {} {\bibfield  {journal} {\bibinfo  {journal} {Phys. Rev. B}\ }\textbf {\bibinfo {volume} {89}},\ \bibinfo {pages} {115401} (\bibinfo {year} {2014})}\BibitemShut {NoStop}%
\bibitem [{\citenamefont {Premakumar}\ \emph {et~al.}(2017)\citenamefont {Premakumar}, \citenamefont {Vavilov},\ and\ \citenamefont {Joynt}}]{premakumar2017evanescent}%
  \BibitemOpen
  \bibfield  {author} {\bibinfo {author} {\bibfnamefont {V.~N.}\ \bibnamefont {Premakumar}}, \bibinfo {author} {\bibfnamefont {M.~G.}\ \bibnamefont {Vavilov}},\ and\ \bibinfo {author} {\bibfnamefont {R.}~\bibnamefont {Joynt}},\ }\bibfield  {title} {\bibinfo {title} {Evanescent-wave johnson noise in small devices},\ }\href@noop {} {\bibfield  {journal} {\bibinfo  {journal} {Quantum Science and Technology}\ }\textbf {\bibinfo {volume} {3}},\ \bibinfo {pages} {015001} (\bibinfo {year} {2017})}\BibitemShut {NoStop}%
\bibitem [{\citenamefont {Tenberg}\ \emph {et~al.}(2019)\citenamefont {Tenberg}, \citenamefont {Asaad}, \citenamefont {Madzik}, \citenamefont {Johnson}, \citenamefont {Joecker}, \citenamefont {Laucht}, \citenamefont {Hudson}, \citenamefont {Itoh}, \citenamefont {Jakob}, \citenamefont {Johnson}, \citenamefont {Jamieson}, \citenamefont {McCallum}, \citenamefont {Dzurak}, \citenamefont {Joynt},\ and\ \citenamefont {Morello}}]{Tenberg2019}%
  \BibitemOpen
  \bibfield  {author} {\bibinfo {author} {\bibfnamefont {S.}~\bibnamefont {Tenberg}}, \bibinfo {author} {\bibfnamefont {S.}~\bibnamefont {Asaad}}, \bibinfo {author} {\bibfnamefont {M.}~\bibnamefont {Madzik}}, \bibinfo {author} {\bibfnamefont {M.}~\bibnamefont {Johnson}}, \bibinfo {author} {\bibfnamefont {B.}~\bibnamefont {Joecker}}, \bibinfo {author} {\bibfnamefont {A.}~\bibnamefont {Laucht}}, \bibinfo {author} {\bibfnamefont {F.}~\bibnamefont {Hudson}}, \bibinfo {author} {\bibfnamefont {K.~M.}\ \bibnamefont {Itoh}}, \bibinfo {author} {\bibfnamefont {A.~M.}\ \bibnamefont {Jakob}}, \bibinfo {author} {\bibfnamefont {B.~C.}\ \bibnamefont {Johnson}}, \bibinfo {author} {\bibfnamefont {D.~N.}\ \bibnamefont {Jamieson}}, \bibinfo {author} {\bibfnamefont {J.~C.}\ \bibnamefont {McCallum}}, \bibinfo {author} {\bibfnamefont {A.~S.}\ \bibnamefont {Dzurak}}, \bibinfo {author} {\bibfnamefont {R.}~\bibnamefont {Joynt}},\ and\ \bibinfo {author} {\bibfnamefont {A.}~\bibnamefont {Morello}},\ }\bibfield  {title} {\bibinfo
  {title} {Electron spin relaxation of single phosphorus donors in metal-oxide-semiconductor nanoscale devices},\ }\href@noop {} {\bibfield  {journal} {\bibinfo  {journal} {Phys. Rev. B}\ }\textbf {\bibinfo {volume} {99}},\ \bibinfo {pages} {205306} (\bibinfo {year} {2019})}\BibitemShut {NoStop}%
\bibitem [{\citenamefont {Undseth}\ \emph {et~al.}(2023)\citenamefont {Undseth}, \citenamefont {Pietx-Casas}, \citenamefont {Raymenants}, \citenamefont {Mehmandoost}, \citenamefont {M{\k{a}}dzik}, \citenamefont {Philips}, \citenamefont {De~Snoo}, \citenamefont {Michalak}, \citenamefont {Amitonov}, \citenamefont {Tryputen} \emph {et~al.}}]{undseth2023hotter}%
  \BibitemOpen
  \bibfield  {author} {\bibinfo {author} {\bibfnamefont {B.}~\bibnamefont {Undseth}}, \bibinfo {author} {\bibfnamefont {O.}~\bibnamefont {Pietx-Casas}}, \bibinfo {author} {\bibfnamefont {E.}~\bibnamefont {Raymenants}}, \bibinfo {author} {\bibfnamefont {M.}~\bibnamefont {Mehmandoost}}, \bibinfo {author} {\bibfnamefont {M.~T.}\ \bibnamefont {M{\k{a}}dzik}}, \bibinfo {author} {\bibfnamefont {S.~G.}\ \bibnamefont {Philips}}, \bibinfo {author} {\bibfnamefont {S.~L.}\ \bibnamefont {De~Snoo}}, \bibinfo {author} {\bibfnamefont {D.~J.}\ \bibnamefont {Michalak}}, \bibinfo {author} {\bibfnamefont {S.~V.}\ \bibnamefont {Amitonov}}, \bibinfo {author} {\bibfnamefont {L.}~\bibnamefont {Tryputen}}, \emph {et~al.},\ }\bibfield  {title} {\bibinfo {title} {Hotter is easier: unexpected temperature dependence of spin qubit frequencies},\ }\href@noop {} {\bibfield  {journal} {\bibinfo  {journal} {Physical Review X}\ }\textbf {\bibinfo {volume} {13}},\ \bibinfo {pages} {041015} (\bibinfo {year} {2023})}\BibitemShut {NoStop}%
\bibitem [{\citenamefont {Choi}\ and\ \citenamefont {Joynt}(2024)}]{choi2024interacting}%
  \BibitemOpen
  \bibfield  {author} {\bibinfo {author} {\bibfnamefont {Y.}~\bibnamefont {Choi}}\ and\ \bibinfo {author} {\bibfnamefont {R.}~\bibnamefont {Joynt}},\ }\bibfield  {title} {\bibinfo {title} {Interacting random-field dipole defect model for heating in semiconductor-based qubit devices},\ }\href@noop {} {\bibfield  {journal} {\bibinfo  {journal} {Physical Review Research}\ }\textbf {\bibinfo {volume} {6}},\ \bibinfo {pages} {013168} (\bibinfo {year} {2024})}\BibitemShut {NoStop}%
\bibitem [{\citenamefont {Amoretti}(2023)}]{Amoretti_2023}%
  \BibitemOpen
  \bibfield  {author} {\bibinfo {author} {\bibfnamefont {A.}~\bibnamefont {Amoretti}},\ }\bibfield  {title} {\bibinfo {title} {Superconductors in strong electric fields: Quantum electrodynamics meets superconductivity},\ }\href@noop {} {\bibfield  {journal} {\bibinfo  {journal} {Journal of Physics: Conference Series}\ }\textbf {\bibinfo {volume} {2531}},\ \bibinfo {pages} {012001} (\bibinfo {year} {2023})}\BibitemShut {NoStop}%
\bibitem [{\citenamefont {Zaccone}\ and\ \citenamefont {Fomin}(2024)}]{zaccone2024theory}%
  \BibitemOpen
  \bibfield  {author} {\bibinfo {author} {\bibfnamefont {A.}~\bibnamefont {Zaccone}}\ and\ \bibinfo {author} {\bibfnamefont {V.~M.}\ \bibnamefont {Fomin}},\ }\bibfield  {title} {\bibinfo {title} {Theory of superconductivity in thin films under an external electric field},\ }\href@noop {} {\bibfield  {journal} {\bibinfo  {journal} {Physical Review B}\ }\textbf {\bibinfo {volume} {109}},\ \bibinfo {pages} {144520} (\bibinfo {year} {2024})}\BibitemShut {NoStop}%
\bibitem [{\citenamefont {Tosato}\ \emph {et~al.}(2023)\citenamefont {Tosato}, \citenamefont {Levajac}, \citenamefont {Wang}, \citenamefont {Boor}, \citenamefont {Borsoi}, \citenamefont {Botifoll}, \citenamefont {Borja}, \citenamefont {Mart{\'\i}-S{\'a}nchez}, \citenamefont {Arbiol}, \citenamefont {Sammak} \emph {et~al.}}]{tosato2023hard}%
  \BibitemOpen
  \bibfield  {author} {\bibinfo {author} {\bibfnamefont {A.}~\bibnamefont {Tosato}}, \bibinfo {author} {\bibfnamefont {V.}~\bibnamefont {Levajac}}, \bibinfo {author} {\bibfnamefont {J.-Y.}\ \bibnamefont {Wang}}, \bibinfo {author} {\bibfnamefont {C.~J.}\ \bibnamefont {Boor}}, \bibinfo {author} {\bibfnamefont {F.}~\bibnamefont {Borsoi}}, \bibinfo {author} {\bibfnamefont {M.}~\bibnamefont {Botifoll}}, \bibinfo {author} {\bibfnamefont {C.~N.}\ \bibnamefont {Borja}}, \bibinfo {author} {\bibfnamefont {S.}~\bibnamefont {Mart{\'\i}-S{\'a}nchez}}, \bibinfo {author} {\bibfnamefont {J.}~\bibnamefont {Arbiol}}, \bibinfo {author} {\bibfnamefont {A.}~\bibnamefont {Sammak}}, \emph {et~al.},\ }\bibfield  {title} {\bibinfo {title} {Hard superconducting gap in germanium},\ }\href@noop {} {\bibfield  {journal} {\bibinfo  {journal} {Communications Materials}\ }\textbf {\bibinfo {volume} {4}},\ \bibinfo {pages} {23} (\bibinfo {year} {2023})}\BibitemShut {NoStop}%
\bibitem [{\citenamefont {Mohanty}(2002)}]{mohanty2002decoherent}%
  \BibitemOpen
  \bibfield  {author} {\bibinfo {author} {\bibfnamefont {P.}~\bibnamefont {Mohanty}},\ }\bibfield  {title} {\bibinfo {title} {Of decoherent electrons and disordered conductors},\ }\href@noop {} {\bibfield  {journal} {\bibinfo  {journal} {Complexity from Microscopic to Macroscopic Scales: Coherence and Large Deviations}\ ,\ \bibinfo {pages} {49}} (\bibinfo {year} {2002})}\BibitemShut {NoStop}%
\bibitem [{\citenamefont {Yang}\ \emph {et~al.}(2016)\citenamefont {Yang}, \citenamefont {Hsu}, \citenamefont {Stano}, \citenamefont {Klinovaja},\ and\ \citenamefont {Loss}}]{yang2016long}%
  \BibitemOpen
  \bibfield  {author} {\bibinfo {author} {\bibfnamefont {G.}~\bibnamefont {Yang}}, \bibinfo {author} {\bibfnamefont {C.-H.}\ \bibnamefont {Hsu}}, \bibinfo {author} {\bibfnamefont {P.}~\bibnamefont {Stano}}, \bibinfo {author} {\bibfnamefont {J.}~\bibnamefont {Klinovaja}},\ and\ \bibinfo {author} {\bibfnamefont {D.}~\bibnamefont {Loss}},\ }\bibfield  {title} {\bibinfo {title} {Long-distance entanglement of spin qubits via quantum hall edge states},\ }\href@noop {} {\bibfield  {journal} {\bibinfo  {journal} {Physical Review B}\ }\textbf {\bibinfo {volume} {93}},\ \bibinfo {pages} {075301} (\bibinfo {year} {2016})}\BibitemShut {NoStop}%
\bibitem [{\citenamefont {Tanamoto}\ and\ \citenamefont {Ono}(2021)}]{tanamoto2021compact}%
  \BibitemOpen
  \bibfield  {author} {\bibinfo {author} {\bibfnamefont {T.}~\bibnamefont {Tanamoto}}\ and\ \bibinfo {author} {\bibfnamefont {K.}~\bibnamefont {Ono}},\ }\bibfield  {title} {\bibinfo {title} {Compact spin qubits using the common gate structure of fin field-effect transistors},\ }\href@noop {} {\bibfield  {journal} {\bibinfo  {journal} {AIP Advances}\ }\textbf {\bibinfo {volume} {11}} (\bibinfo {year} {2021})}\BibitemShut {NoStop}%
\bibitem [{\citenamefont {K{\k{e}}pa}\ \emph {et~al.}(2023{\natexlab{a}})\citenamefont {K{\k{e}}pa}, \citenamefont {Cywi{\'n}ski},\ and\ \citenamefont {Krzywda}}]{kkepa2023correlations}%
  \BibitemOpen
  \bibfield  {author} {\bibinfo {author} {\bibfnamefont {M.}~\bibnamefont {K{\k{e}}pa}}, \bibinfo {author} {\bibfnamefont {{\L}.}~\bibnamefont {Cywi{\'n}ski}},\ and\ \bibinfo {author} {\bibfnamefont {J.~A.}\ \bibnamefont {Krzywda}},\ }\bibfield  {title} {\bibinfo {title} {Correlations of spin splitting and orbital fluctuations due to 1/f charge noise in the {S}i/{S}i{G}e quantum dot},\ }\href@noop {} {\bibfield  {journal} {\bibinfo  {journal} {Applied Physics Letters}\ }\textbf {\bibinfo {volume} {123}},\ \bibinfo {pages} {034003} (\bibinfo {year} {2023}{\natexlab{a}})}\BibitemShut {NoStop}%
\bibitem [{\citenamefont {K{\k{e}}pa}\ \emph {et~al.}(2023{\natexlab{b}})\citenamefont {K{\k{e}}pa}, \citenamefont {Cywi{\'n}ski},\ and\ \citenamefont {Krzywda}}]{kkepa2023simulation}%
  \BibitemOpen
  \bibfield  {author} {\bibinfo {author} {\bibfnamefont {M.}~\bibnamefont {K{\k{e}}pa}}, \bibinfo {author} {\bibfnamefont {{\L}.}~\bibnamefont {Cywi{\'n}ski}},\ and\ \bibinfo {author} {\bibfnamefont {J.~A.}\ \bibnamefont {Krzywda}},\ }\bibfield  {title} {\bibinfo {title} {{Simulation of 1/f charge noise affecting a quantum dot in a {S}i/{S}i{G}e structure}},\ }\href@noop {} {\bibfield  {journal} {\bibinfo  {journal} {Applied Physics Letters}\ }\textbf {\bibinfo {volume} {123}},\ \bibinfo {pages} {034005} (\bibinfo {year} {2023}{\natexlab{b}})}\BibitemShut {NoStop}%
\bibitem [{\citenamefont {Yoneda}\ \emph {et~al.}(2023)\citenamefont {Yoneda}, \citenamefont {Rojas-Arias}, \citenamefont {Stano}, \citenamefont {Takeda}, \citenamefont {Noiri}, \citenamefont {Nakajima}, \citenamefont {Loss},\ and\ \citenamefont {Tarucha}}]{yoneda2023noise}%
  \BibitemOpen
  \bibfield  {author} {\bibinfo {author} {\bibfnamefont {J.}~\bibnamefont {Yoneda}}, \bibinfo {author} {\bibfnamefont {J.}~\bibnamefont {Rojas-Arias}}, \bibinfo {author} {\bibfnamefont {P.}~\bibnamefont {Stano}}, \bibinfo {author} {\bibfnamefont {K.}~\bibnamefont {Takeda}}, \bibinfo {author} {\bibfnamefont {A.}~\bibnamefont {Noiri}}, \bibinfo {author} {\bibfnamefont {T.}~\bibnamefont {Nakajima}}, \bibinfo {author} {\bibfnamefont {D.}~\bibnamefont {Loss}},\ and\ \bibinfo {author} {\bibfnamefont {S.}~\bibnamefont {Tarucha}},\ }\bibfield  {title} {\bibinfo {title} {Noise-correlation spectrum for a pair of spin qubits in silicon},\ }\href@noop {} {\bibfield  {journal} {\bibinfo  {journal} {Nature Physics}\ }\textbf {\bibinfo {volume} {19}},\ \bibinfo {pages} {1793} (\bibinfo {year} {2023})}\BibitemShut {NoStop}%
\bibitem [{\citenamefont {Machlup}(1954)}]{machlup1954noise}%
  \BibitemOpen
  \bibfield  {author} {\bibinfo {author} {\bibfnamefont {S.}~\bibnamefont {Machlup}},\ }\bibfield  {title} {\bibinfo {title} {Noise in semiconductors: spectrum of a two-parameter random signal},\ }\href@noop {} {\bibfield  {journal} {\bibinfo  {journal} {Journal of Applied Physics}\ }\textbf {\bibinfo {volume} {25}},\ \bibinfo {pages} {341} (\bibinfo {year} {1954})}\BibitemShut {NoStop}%
\bibitem [{\citenamefont {Mehmandoost}\ and\ \citenamefont {Dobrovitski}(2024)}]{mehmandoost2024decoherence}%
  \BibitemOpen
  \bibfield  {author} {\bibinfo {author} {\bibfnamefont {M.}~\bibnamefont {Mehmandoost}}\ and\ \bibinfo {author} {\bibfnamefont {V.}~\bibnamefont {Dobrovitski}},\ }\bibfield  {title} {\bibinfo {title} {Decoherence induced by a sparse bath of two-level fluctuators: peculiar features of $1/f $ noise in high-quality qubits},\ }\href@noop {} {\bibfield  {journal} {\bibinfo  {journal} {arXiv preprint arXiv:2404.18659}\ } (\bibinfo {year} {2024})}\BibitemShut {NoStop}%
\bibitem [{\citenamefont {Kawakami}\ \emph {et~al.}(2014)\citenamefont {Kawakami}, \citenamefont {Scarlino}, \citenamefont {Ward}, \citenamefont {Braakman}, \citenamefont {Savage}, \citenamefont {Lagally}, \citenamefont {Friesen}, \citenamefont {Coppersmith}, \citenamefont {Eriksson},\ and\ \citenamefont {Vandersypen}}]{Kawakami:2014p666}%
  \BibitemOpen
  \bibfield  {author} {\bibinfo {author} {\bibfnamefont {E.}~\bibnamefont {Kawakami}}, \bibinfo {author} {\bibfnamefont {P.}~\bibnamefont {Scarlino}}, \bibinfo {author} {\bibfnamefont {D.~R.}\ \bibnamefont {Ward}}, \bibinfo {author} {\bibfnamefont {F.~R.}\ \bibnamefont {Braakman}}, \bibinfo {author} {\bibfnamefont {D.~E.}\ \bibnamefont {Savage}}, \bibinfo {author} {\bibfnamefont {M.~G.}\ \bibnamefont {Lagally}}, \bibinfo {author} {\bibfnamefont {M.}~\bibnamefont {Friesen}}, \bibinfo {author} {\bibfnamefont {S.~N.}\ \bibnamefont {Coppersmith}}, \bibinfo {author} {\bibfnamefont {M.~A.}\ \bibnamefont {Eriksson}},\ and\ \bibinfo {author} {\bibfnamefont {L.~M.~K.}\ \bibnamefont {Vandersypen}},\ }\bibfield  {title} {\bibinfo {title} {Electrical control of a long-lived spin qubit in a {Si/SiGe} quantum dot},\ }\href@noop {} {\bibfield  {journal} {\bibinfo  {journal} {Nat. Nanotechnol.}\ }\textbf {\bibinfo {volume} {9}},\ \bibinfo {pages} {666} (\bibinfo {year} {2014})}\BibitemShut {NoStop}%
\end{thebibliography}%

\end{document}